\documentclass[11pt]{article}
\pdfoutput=1 

\usepackage{jheppub} 

\usepackage[T1]{fontenc} 

\usepackage{bm}
\usepackage{bigdelim}
\usepackage[table,dvipsnames]{xcolor}
  \def\red{\textcolor{red}}
  \def\blue{\textcolor{blue}}
  \def\magenta{\textcolor{magenta}}

\usepackage{footnote}

\def\alphas{\alpha_{\rm s}}
\def\Nc{N_{\rm c}}
\def\CA{C_{\rm A}}
\def\CF{C_{\rm F}}
\def\qhatA{\hat q_{\rm A}}
\def\Re{\operatorname{Re}}
\def\gammaE{\gamma_{\rm\scriptscriptstyle E}}
\def\eps{\epsilon}

\def\b{{\bm b}}
\def\p{{\bm p}}
\def\q{{\bm q}}
\def\x{{\bm x}}
\def\B{{\bm B}}
\def\P{{\bm P}}
\def\B{{\bm B}}
\def\grad{{\bm\nabla}}

\def\LMW{{\scriptscriptstyle\rm LMW}}

\def\CFas{\CF\kern0.6pt\alphas}
\def\CFgg{\CF\kern0.4pt g^2}

\title{\boldmath Universality (beyond leading log) of soft radiative corrections
   to $\hat q$ in $p_\perp$ broadening and energy loss}

\author{Peter Arnold}
\affiliation{Department of Physics, University of Virginia,
  P.O.\ Box 400714, 
  Charlottesville, VA 22904, U.S.A.}

\emailAdd{parnold@virginia.edu}

\abstract{
  It has been known for many years that soft radiation can give potentially
  large double-logarithm corrections to $p_\perp$ broadening of a high-energy
  particle traveling through QCD matter, but that this soft radiation
  correction can be absorbed into an effective value $\hat q_{\rm eff}$
  for the medium
  $p_\perp$-broadening parameter $\hat q$.  Here ``soft'' means high
  energy compared to medium scales but soft compared to the original
  high-energy particle traveling through the medium.
  A similar situation arises in the case of soft corrections to
  hard splitting of a high-energy particle, such as hard $g{\to}gg$,
  where double logarithms can also be absorbed using the same effective
  $\hat q_{\rm eff}$.  In this paper, I study whether the same holds true for
  potentially large, subleading, {\it single}-logarthim corrections.
  The correspondence is more indirect for single logarithms,
  but I show (in the large-$\Nc$ limit)
  that single logarithms from soft radiation in the
  case of $p_\perp$ broadening also determine the single logarithms
  from soft radiative corrections to hard $g{\to}gg$ splitting.
  Along the way, there is an interesting variation of the original BDMPS-Z
  calculation of splitting rates in the $\hat q$ approximation.
  I also discuss how, for soft-radiative corrections to
  hard splitting processes, there are two different types
  of ``$\hat q_{\rm eff}$'' that come into play, which differ by
  ``$i\pi$'' terms that multiply single logarithms.
}

\begin{document} 
\maketitle
\flushbottom

\newpage


\section{Introduction}
\label{sec:intro}

High-energy partons traveling through hot or cold QCD matter receive
random transverse momentum kicks from multiple small-angle scattering with the
medium.  The typical total transverse momentum change $p_\perp$ after
traveling through a length $L$ of medium involving many such interactions
is parametrized as
\begin {equation}
   \langle p_\perp^2 \rangle = \hat q L ,
\label {eq:broaden}
\end {equation}
where $\hat q$ is determined by characteristics of the medium.
$\hat q$ also appears in formulas for high-energy parton splitting
rates in the medium.
For example, formalism developed by
Baier, Dokshitzer, Mueller, Peign\'e, and Schiff
\cite{BDMPS1,BDMPS2,BDMPS3,BDMS}
and Zakharov \cite{Zakharov1,Zakharov2,Zakharov3}
(BDMPS-Z) gives (in appropriate limits) the in-medium
gluon splitting rate%
\footnote{
  It's difficult to figure out whom to reference for the first appearance
  of (\ref{eq:LOrate0}).  BDMS \cite{BDMS} give the $q{\to}qg$ formula
  in their eq.\ (42b) [with the relevant limit here being the infinite
  volume limit $\tau_0 \to \infty$ for their time $\tau_0$].  They then
  discuss elements of the $g{\to}gg$ case after that but don't quite
  give an explicit formula for the entire rate.  (They are not explicit
  about the formula for $\omega_0$.)
  Zakharov makes a
  few general statements about the $g{\to}gg$ case
  after eq.\ (75) of ref.\ \cite{Zakharov3}.
  As an example from ten years later, the explicit formula is
  given by eqs.\ (2.26) and (4.6) of ref.\ \cite{simple} in the
  case where $s$ represents a gluon.
}
\begin {equation}
  \frac{d\Gamma}{dx}
  = \frac{\alphas P_{g\to gg}(x)}{2\pi}
    \sqrt{ \frac{(1{-}x{+}x^2) \qhatA}{x(1{-}x)E} }
\label {eq:LOrate0}
\end {equation}
for $g \to gg$ with energies $E \to xE,(1{-}x)E$.
(The subscript of $\qhatA$ indicates the $\hat q$ appropriate
for the adjoint color representation, i.e. for gluons, and
$\CA{=}\Nc$ is the adjoint-representation quadratic Casimir.)

\begin {figure}[t]
\begin {center}
  \resizebox{0.95\linewidth}{!}{
  \begin{picture}(470,80)(0,0)
    \put(40,25){\includegraphics[scale=0.4]{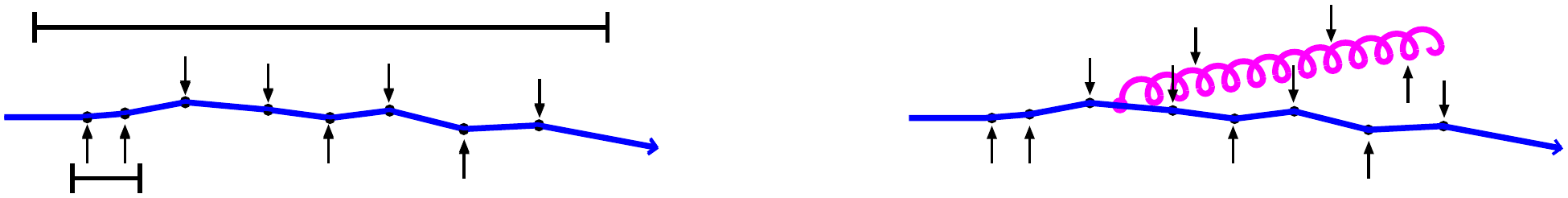}}
    \put(0,42){\blue{large $E$}}
    \put(200,35){${\blue{\langle p_\perp^2 \rangle}} {=} \hat q L$}
    \put(80,70){distance $L \gg \tau_0$}
    \put(60,22){$\tau_0$}
    \put(110,2){(a)}
    \put(390,60){\magenta{soft $\omega_y {=} yE \ll E$}}
    \put(414,35){
        ${\blue{\langle p_\perp^2 \rangle}} {=} {\magenta{\hat q_{\rm eff}}} L$
    }
    \put(327,2){(b)}
  \end{picture}
  } 
  \caption{
     \label {fig:LMWqhat}
     (a) A cartoon of transverse momentum broadening as a high-energy
     particle traverses a QCD medium.  The arrows represent small
     transverse momentum kicks from the medium due to small-angle
     elastic scattering with the medium.
     (b) The same, but here some transverse momentum is also carried
     away by a relatively soft (but still high-energy) bremsstrahlung.
  }
\end {center}
\end {figure}

\begin {figure}[t]
\begin {center}
  \resizebox{0.95\linewidth}{!}{
  \begin{picture}(470,80)(0,0)
    \put(70,25){\includegraphics[scale=0.4]{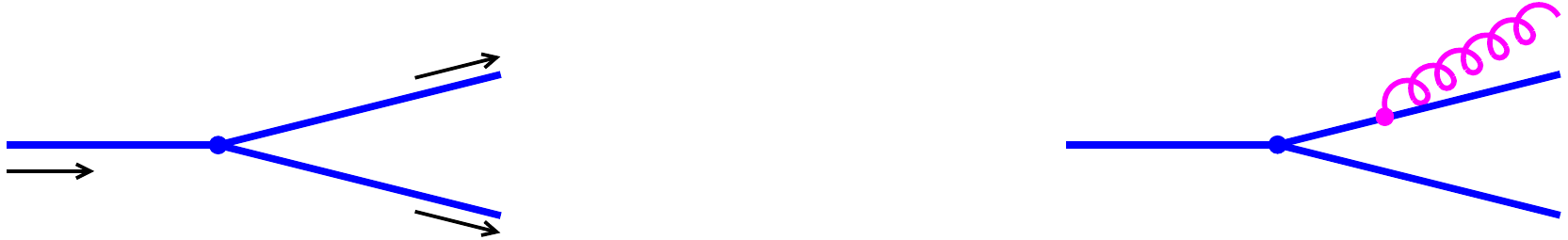}}
    \put(32,42){\blue{large $E$}}
    \put(176,56){\blue{hard $xE$}}
    \put(176,28){\blue{hard $(1{-}x)E$}}
    \put(110,2){(a)}
    \put(390,72){\magenta{soft $\omega_y {=} yE \ll E$}}
    \put(327,2){(b)}
  \end{picture}
  } 
  \caption{
     \label {fig:LOsplit+SOFT}
     (a) A hard, underlying splitting process (bremsstrahlung or pair
     production) with energies $E \to xE,(1{-}x)E$.
     (b) One example of adding an additional, softer gluon bremsstrahlung,
     which should be understood as occurring during the formation time of
     the underlying, harder splitting process (a).
  }
\end {center}
\end {figure}

In the original picture (fig.\ \ref{fig:LMWqhat}a)
of momentum broadening motivating
(\ref{eq:broaden}), $\hat q$ is determined by small-angle elastic
scattering rates of high-energy particles scattering from the medium.
Liou, Mueller, and Wu (LMW) \cite{LMW} realized that soft gluon
radiation accompanying elastic scatterings (fig.\ \ref{fig:LMWqhat}b) can also
carry away transverse momentum and so change $p_\perp$ in an
important way.  Formally, this effect is suppressed by
a power of $\alphas$ but is enhanced by what can be a potentially
large double logarithm.  They found that such logarithms could
be absorbed into an effective value
$\hat q_{\rm eff} = \hat q + \delta q$ of $\hat q$.
To leading-log order, the soft radiation effects are accounted for by
\begin {equation}
  \delta \hat q
  = \frac{\CA \alphas}{2\pi} \ln^2\Bigl( \frac{L}{\tau_0} \Bigr) \, \hat q ,
\label {eq:LMWlogsqr}
\end {equation}
where $\tau_0$ is the scale of the mean-free path for
typical small-angle elastic
scattering in the medium.
They also worked out how to resum the
effects of multiple soft gluon bremsstrahlung to leading-log order.
Later, various authors \cite{Blaizot,Iancu,Wu} investigated
similar effects for in-medium splitting rates such as
(\ref{eq:LOrate0}).  That is, what would be the effect on
$q \to qg$ or $g \to gg$ (fig.\ \ref{fig:LOsplit+SOFT}a)
of having an {\it additional}\/ softer bremsstrahlung
(fig.\ \ref{fig:LOsplit+SOFT}b)
occur during an underlying, harder in-medium splitting process.
They again found a double logarithm.  Moreover, they
found that the effect was completely accounted for, at leading-log
order, simply by making the same modification (\ref{eq:LMWlogsqr})
to the $\hat q$ appearing in splitting rates such as (\ref{eq:LOrate0}).
So, there are important soft radiative corrections, but they are
{\it universal}\/ in that they can be absorbed into $\hat q$ in a way that
is independent of whether your interest is $p_\perp$ broadening
or splitting rates.

In the context of $p_\perp$ broadening,
LMW also computed the {\it single}-log correction, sub-leading to
the double-log correction (\ref{eq:LMWlogsqr}).
In this paper, I examine whether the single-log correction is
also universal.  At first sight, it may not seem to be.
Ref.\ \cite{logs} recently extracted, in the large-$\Nc$ limit,
the soft single-log
corrections to the hard splitting rate (\ref{eq:LOrate0})
from more-general results for $g \to ggg$
(e.g.\ fig.\ \ref{fig:LOsplit+SOFT}b with nothing soft).
As I will review,
the coefficient of that single log is a slightly complicated function
of the energy fraction $x$ of the daughter of the hard splitting process
(fig.\ \ref{fig:LOsplit+SOFT}a), which has no analog in the discussion
of $p_\perp$ broadening (fig.\ \ref{fig:LMWqhat}a).
Nonetheless, we will see that there {\it is} a connection.
I will show that the slightly complicated soft single-log correction to
hard splitting can be exactly reproduced from the simpler LMW result
for the soft single-log correction to $p_\perp$ broadening.
We will see that this requires re-doing the BDMPS-Z calculation of
(\ref{eq:LOrate0}) in a more general way that allows
incorporation of the LMW result.

That calculation will verify, in a highly non-trivial example,
that the soft single-log corrections to $\hat q$ are universal
(and that they completely account for all soft single-log corrections
to hard splitting).  This will also be an important cross-check of the
more general calculation of the {\it non}-soft overlapping splitting
$g \to gg \to ggg$ in refs.\ \cite{2brem,seq,dimreg,QEDnf,qcd}.

It would be nice to also have a relation between the single logs
that is not embedded in the cogs of a re-derivation of the BDMP-Z
formula.
Inspired by that equivalence, I later find a way to algebraically rewrite the
formula for the
previously known single-log correction \cite{logs} to hard splitting
in ways that more directly connect to the LMW result for
$p_\perp$ broadening.  That version will be a re-writing, not a
re-derivation, of the single logs.  But it will isolate more clearly the
physics of the single log result and may be useful in applications.


\subsection*{Outline}

In the next section, I review the phase space for soft emission that
gives rise to double logarithms, and I give some important caveats about
exactly what will be checked in this paper regarding single logarithms.
In section \ref{sec:known}, I then quote the already-known results
for single logarithms from soft radiative corrections to
$p_\perp$ broadening and to hard $g{\to}gg$ splitting.
I also give a short, qualitative review of the formalism underlying LMW's
calculation for $p_\perp$ broadening
and explain why one cannot instantly apply their
$\hat q_{\rm eff}$ result to the usual rate formula for hard $g{\to}gg$
splitting.  Section \ref{sec:BDMPSZ} briefly
reviews the BDMPS-Z based derivation
of the hard splitting rate (\ref{eq:LOrate0}) as preparation for
modifying that derivation in section \ref{sec:calculation} to
properly incorporate the $\hat q_{\rm eff}$ from $p_\perp$ broadening.
Section \ref{sec:calculation} will successfully
reproduce the single logarithm previously extracted in ref.\ \cite{logs},
but using here a {\it much} simpler calculation that {\it assumes}
universality of soft radiative
corrections to $\hat q$.
We will also see that, at the level of soft radiative corrections,
there is a difference between a $\hat q_{\rm eff}$ involving (i) one particle
in the amplitude and one in the complex conjugate amplitude vs.\
(ii) two particles in the amplitude.  One gets both types of $\hat q$'s
appearing in splitting calculations; to my knowledge, a difference
between (i) and (ii) in this context has not previously been demonstrated.%
\footnote{
  \label{foot:simon}
  Caron-Huot speculated on this possibility at the end of section 3.1
  of ref.\ \cite{simon}.  He also speculated in private
  correspondence (2018) that there would be a difference corresponding
  to $i\pi$ terms, which is indeed the type of difference I now find
  in (\ref{eq:bbvsLMW}).
  As motivation, he pointed me to the $i\pi$ terms in eq.\ (51) of
  ref.\ \cite{iPi}, which represents an NLO dipole-dipole scattering
  amplitude in vacuum computed from light-like Wilson lines.
}
In section \ref{sec:rewrites}, I show a variety of ways of writing
the final result for soft corrections to hard $g{\to}gg$ splitting
in terms of the result for $\hat q_{\rm eff}$ from
$p_\perp$ broadening, along with some qualitative explanation of the
results.  The most compact formulation is
presented in section \ref{sec:conclusion}, where I give my conclusions.


\section {Caveats and Assumptions}

\subsection {The double log region}

Throughout this paper, I will draw diagrams for contributions to
splitting rates using the conventions of ref.\ \cite{2brem}, which
are adapted from Zakharov's description of splitting rates
\cite{Zakharov1,Zakharov2,Zakharov3}.
In fig.\ \ref{fig:split}a,
the blue factor represents the
calculation of the splitting amplitude in
(lightcone) time-ordered perturbation theory.
The red factor is the conjugate amplitude.
Fig.\ \ref{fig:split}b shows the same time-ordered
contribution to the rate, depicted
by sewing together the amplitude and conjugate
amplitude.
I then follow Zakharov's picture of re-interpreting the right-hand
diagrams as three particles propagating forward in time.
Only the high-energy particle lines are shown in these diagrams:
the lines implicitly interact with the medium as they propagate, and there
is an implicit average of the rate over the randomness of the medium.

\begin{savenotes}
\begin {figure}[t]
\begin {center}
  \includegraphics[scale=0.6]{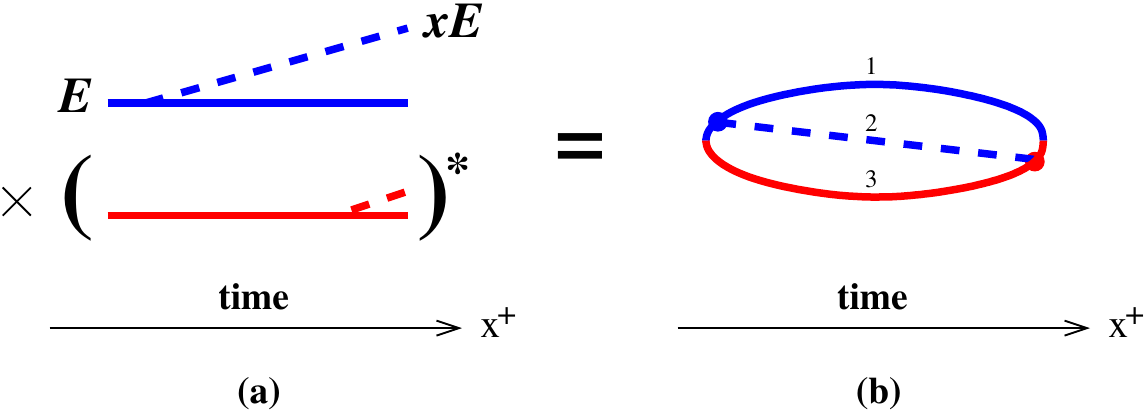}
  \caption{
     \label {fig:split}
     (a) A time-ordered contribution to the rate for single splitting,
     such as $g \to gg$ of high-energy gluons, with
     amplitude in blue and conjugate amplitude in red.
     (b) A single diagram representing this contribution to the rate.
     In both cases, all lines implicitly interact
     with the medium.  We need not follow
     particles after the emission has occurred in both the amplitude
     and conjugate amplitude because I will consider only the
     $p_\perp$-integrated rate.
     (See, for example, section 4.1 of
     ref.\ \cite{2brem} for a more explicit argument,
     although applied there to a more complicated diagram.)
     Nor need we follow them before
     the first emission because we approximate the initial particle
     as on-shell.
     Only one of the two time orderings that contribute to the
     rate is shown above.
  }
\end {center}
\end {figure}
\end {savenotes}

\begin {figure}[t]
\begin {center}
  \includegraphics[scale=0.6]{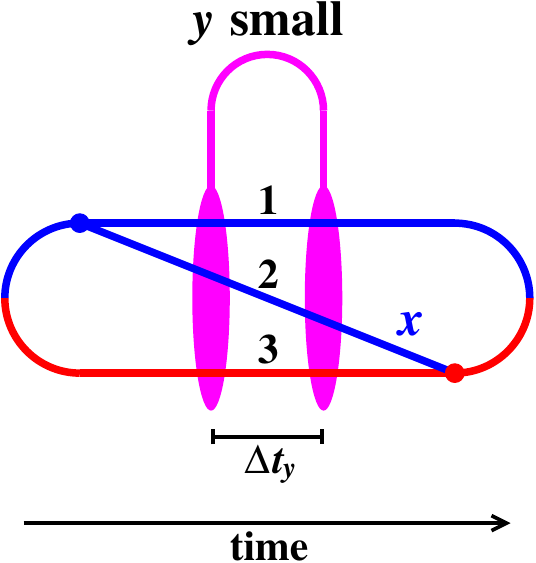}
  \caption{
     \label{fig:DLOGdiags}
     The rate diagrams which (together with their complex conjugates)
     produce the double logarithm \cite{Iancu,Blaizot,Wu}.
     On each end, the relatively-soft $y$ gluon may connect to any
     of the three lines of the underlying splitting process.
     In my analysis, all of the underlying lines will also be gluons.
     The magenta color of the $y$ gluon is used here to indicate that
     it could be colored blue or red depending on whether
     its left-hand end is connected to a blue or red line, respectively.
  }
\end {center}
\end {figure}

Fig.\ \ref{fig:DLOGdiags} depicts a soft radiative correction
(the magenta line) to the underlying splitting process of
fig.\ \ref{fig:split}b.
Throughout this paper, I will refer
to the energy of the initial high-energy particle in the underlying
hard splitting process as $E$ and to the energy
of the soft radiative gluon as
\begin {equation}
   \omega_y \equiv y E ,
\end {equation}
as in fig.\ \ref{fig:LOsplit+SOFT}.
As shown in fig.\ \ref{fig:DLOGdiags},
I define $\Delta t_y$ to be the separation between times
of the $y$ emissions
in the amplitude and conjugate amplitude, which is to be integrated over.

For the underlying hard splitting process $E \to xE, (1{-}x)E$, it will
be convenient for qualitative, parametric discussions
to take $xE$ to be the smaller of the two daughters.
The eventual derivations and results, however, will be symmetric with
respect to $x {\to} 1{-}x$.
I will {\it not}\/ assume $x \ll 1$ (though that case is also
covered by the analysis provided $y \ll x$).
It's sometimes convenient to also
write the $x$ daughter energy as $\omega_x$:
\begin {equation}
   \omega_x \equiv x E .
\end {equation}
In my analysis, ``soft'' $y$ means soft compared to $x$ but still
high-energy compared to medium scales:
\begin {equation}
  \mbox{(medium scale)} \ll \omega_y \ll \omega_x .
\end {equation}
(For a quark-gluon plasma, ``medium scale'' above just means the
temperature $T$.)  Parametrically, I will refer to the typical
duration of the splitting
process as the $x$ formation time
\begin {equation}
   t_{\rm form}(x) \sim \sqrt{ \frac{\omega_x\vphantom{E}}{\hat q} }
   = \sqrt{ \frac{x E}{\hat q} } \,.
\end {equation}
Here and throughout, I will study the simple case where the medium is
homogeneous over the formation time (and corresponding length).

\begin {figure}[t]
\begin {center}
  \begin{picture}(370,330)(0,0)
    \put(130,90){\includegraphics[scale=1.0]{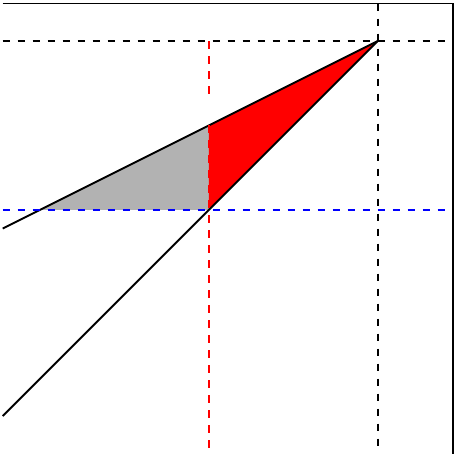}}
    \put(225,318){\scalebox{1.2}{$\bm{\ln{\bm\omega}_y}$}}
    \put(355,200){\rotatebox{-90}{\scalebox{1.2}{$\bm{\ln{\bm\Delta}t_y}$}}}
    \put(8,288){
        $\Delta t \sim t_{\rm form}(x) \sim \sqrt{\omega_x/\hat q}$ }
    \put(81,205){$\color{black}{\Delta t_y\,\sim\,\tau_0}$}
    \put(225,85){\rotatebox{-90}{$\color{black}{
          \omega_y\,\sim\,\hat q \tau_0^{} \, t_{\rm form}(x)
     }$}}
    \put(307,85){\rotatebox{-90}{$\omega_y\,\sim\,\omega_x$}}
    \put(255,223){\rotatebox{45}{$k_{\perp y} \sim k_{\perp x}$}}
    \put(160,222){\rotatebox{27}{
         $\Delta t_y \sim t_{\rm form}(y) \sim \sqrt{\omega_y/\hat q}$ }}
  \end{picture}
  \caption{
     \label {fig:LMWregion}
     The double-log region of the soft gluon ($y$) emission parameter space
     in fig.\ \ref{fig:DLOGdiags} corresponds to all of the shaded regions
     above.  Single logs are determined by the
     behavior at the boundaries of this region.  In this paper, I
     restrict attention to the red-colored region and the single
     log contributions from its boundaries.
  }
\end {center}
\end {figure}

With this notation, the shaded areas of fig.\ \ref{fig:LMWregion}
correspond to the parametric region of $(\omega_y,\Delta t_y)$ that
generates the double log.%
\footnote{
   For small $x$, this parameter region for $(\omega_y,\Delta t_y)$
   is equivalent to that discussed
   in LMW \cite{LMW} in the context of leading-log resummation.
}
If one strictly sticks to the $\hat q$ approximation, the double log
region corresponds to
\begin {equation}
   \frac{\omega_y}{\hat q \, t_{\rm form}(x)} \ll \Delta t_y \ll t_{\rm form}(y) .
\label {eq:tregion}
\end {equation}
The second inequality just says that the time $\Delta t_y$ for the
$y$ emission must fit within the $y$ formation time
$t_{\rm form}(y) \sim \sqrt{\omega_y/\hat q}$.
[For $\Delta t \gg t_{\rm form}(y)$, scattering with the medium decoheres
the $y$ emission process.]
The physical significance of the first inequality is more apparent
by re-expressing (\ref{eq:tregion}) as constraints on transverse
momenta:%
\footnote{
  One way to see the equivalence is to consider that,
  ignoring medium effects,
  emission of the $y$ gluon would be off-shell in energy by
  $\Delta E_y \sim k_{\perp y}^2/2\omega_y$. By the uncertainty
  principle, this can only last a time $\Delta t_y \sim 1/\Delta E_y$
  without some interaction that can put it on shell, and so
  $k_{\perp y}^2 \sim 2\omega_y/\Delta t_y$.  Similarly,
  $k_{\perp x}^2 \sim 2\omega_x/\Delta t_x$, but the dominant
  time scale in the underlying splitting process is
  $\Delta t_x \sim t_{\rm form}(x)$.  Then (\ref{eq:ktregion})
  translates to (\ref{eq:tregion}).
}
\begin {equation}
   k_{\perp x}^2 \gg k_{\perp y}^2 \gg \hat q\, \Delta t_y .
\label {eq:ktregion}
\end {equation}
The first inequality is transverse momentum ordering
$k_{\perp y} \ll k_{\perp x}$ and ensures that the $y$ emission does
not disrupt the underlying $x$-emission process.  In this
language, the second
inequality ensures that the accumulated
transverse momentum transfer from the medium
during the $y$ emission ($\Delta k_{\perp y} \sim \sqrt{\hat q\, \Delta t_y}$\,)
does not disrupt the soft $y$-emission process.

If (\ref{eq:tregion}) were the only constraints, then the
double-log region would cover an infinite area between the two sloped lines in
fig.\ \ref{fig:LMWregion}, which means that the double log would be
infrared divergent.  This divergence is cut off, however, because the
$\hat q$ approximation is a multiple-scattering approximation, and it
becomes senseless for describing the $y$ emission once the time
$\Delta t_y$ for that emission becomes less than or order the
mean-free time $\tau_0$ for small-angle collisions with the medium.
This is the origin of the $\Delta t \gg \tau_0$ constraint on the
double log region in fig.\ \ref{fig:LMWregion}.

Fig.\ 5 is similar to the double-log region discussed by LMW \cite {LMW} for
transverse momentum broadening, except that the duration
$t_{\rm form}(x)$ of the underlying hard splitting process
plays the role here that the length $L$ of medium traversed
plays in the case of transverse momentum broadening.


\subsection {Significant caveats}

Double logs arise from integration over the shaded parametric regions of
fig.\ \ref{fig:LMWregion}.  Sub-leading, single logs are determined by
the behavior of the integration at/near the {\it boundaries} of
the double log region.
The goal of this paper is to first show how to apply LMW's
momentum broadening results to find single log corrections
to splitting processes, and to then {\it verify} that result by
comparison with single logs extracted much more laboriously
\cite{logs}
from generic-$y$ (not specifically small-$y$) results \cite{qcd}
for double splitting $g \to ggg$.  However, those generic-$y$
calculations have so far been performed only in the $\hat q$ approximation,
and so cannot account for single logs coming from the
horizontal lower boundary $\Delta t \sim \tau_0$ in fig.\ \ref{fig:LMWregion}.
I will not attempt to study the physics of the
breakdown of the $\hat q$ approximation.  Instead, in this paper
I will restrict
attention to the double and single logs coming from the red region
of fig.\ \ref{fig:LMWregion}, where the $\hat q$ approximation can
be used at all the boundaries.

The double logs are universal for any value of $\Nc$.
The generic-$y$ formulas of ref.\ \cite{qcd}, and so the
single logs extracted from them, were derived
in the large-$\Nc$ limit.  Similarly, an important step
in this paper will be justified by appeal to the large-$\Nc$
limit.  Like the double logs, the result might be the same
for general $\Nc$, but I do not currently know a way to argue it.


\subsection {Customary caveats}

For the purpose of this paper,
I will treat the original ``bare'' value of $\hat q$
in $\hat q_{\rm eff} = \hat q + \delta \hat q$ as a constant, independent
of energy.
There are caveats and counter-caveats concerning
logarithmic dependence of that approximation
which I will simply ignore in this paper.%
\footnote{
  For example, for fixed-coupling calculations for a weakly-coupled
  medium, the large-$q_\perp$
  Rutherford tail $d\Gamma_{\rm el}/d(q_\perp^2) \propto \alphas^2/q_\perp^4$
  of the elastic scattering cross-section
  causes logarithmic dependence of $\langle q_\perp^2 \rangle$ on
  the upper scale of $q_\perp$ relevant to the process under
  consideration.  On the other hand, including running of $\alphas$
  as $d\Gamma_{\rm el}/d(q_\perp^2) \propto \alphas^2(q_\perp)/q_\perp^4$
  is enough to eventually tame that dependence if the relevant
  upper scale $Q_\perp$ for $q_\perp$ is large enough that
  $\alphas(Q_\perp)$ is small compared to the strength of
  $\alphas$ at medium
  scales.  (See, for example, section VI.B of ref.\ \cite{DeepLPM}, which
  combined earlier observations of refs.\ \cite{BDMPS3} and \cite{Peshier}.)
}

In general, of course, $\alphas$ depends on scale. 
The $\alphas$ associated with a high-energy bremsstrahlung
($\omega_y \gg{}$the medium scale) may be moderately small even if the
medium itself is strongly coupled.
I will formally assume that this ``bremsstrahlung $\alphas$'' is small.
For simplicity, I will also ignore its running
(other than in the motivation for treating it as small).


\section{Known double and single log results}
\label {sec:known}

\subsection {Soft corrections to hard splitting $g{\to}gg$}

Having explained what will be calculated, I can now quote the
single log result found in ref.\ \cite{logs} for $g{\to}gg$.
For small $y$, the differential rate corresponding to the
soft-radiative corrections of fig.\ \ref{fig:DLOGdiags} was found to be
\begin {subequations}
\begin {equation}
   \delta \left[ \frac{d\Gamma}{dx} \right] =
   - \frac{\CA \alphas}{4\pi}
   \left[ \frac{d\Gamma}{dx} \right]_{\rm LO}
      \int_{y \ll x} dy \>
      \frac{ \ln y + \bar s(x) }{y} \,,
\label {eq:syint}
\end {equation}
where $[ d\Gamma/dx ]_{\rm LO}$ is the leading-order splitting
rate (\ref{eq:LOrate0})
and
\begin {equation}
  \bar s(x) =
      - \ln\bigl(16\,x(1{-}x)(1{-}x{+}x^2)\bigr)
      + 2 \, \frac{ \bigl[
                 x^2 \bigl( \ln x - \frac{\pi}{8} \bigr)
                 + (1{-}x)^2 \bigl( \ln(1{-}x) - \frac{\pi}{8} \bigr)
               \bigr] }
             { (1-x+x^2) }
      .
\label {eq:sbar}
\end {equation}
\end {subequations}
The $\Delta t_y$ of
figs.\ \ref{fig:DLOGdiags} and \ref{fig:LMWregion}
has already been integrated over.
In (\ref{eq:syint}), the designation ``$y \ll x$''
for the soft integration region
should be understood as
shorthand for the more general and symmetric condition
that $y \ll \min(x,1{-}x)$.

If one integrates (\ref{eq:syint}) with a small IR cut-off
$y_{\rm cut}{\ll}1$  on $y$, the IR logarithms from (\ref{eq:syint}) become%
\footnote{
  If $x \ll 1$, the single log coefficient (\ref{eq:sbar}) becomes
  $\bar s(x) \simeq - \ln x - 4\ln 2 - \frac{\pi}{4}$.
  In this limit, one might wish to re-organize the classification
  of double vs.\ single logs in (\ref{eq:sycut}) in terms of
  $\ln(y_{\rm cut}/x)$ rather than $\ln y_{\rm cut}$.
  (See section 4.2 of ref.\ \cite{qcd} for
  more discussion.)
  In order to be able to also talk about the case $x{\sim}1$ of an
  underlying {\it hard}\/ $g{\to}gg$ splitting
  (and also
  to symmetrically treat the limits $x \ll 1$ and $1{-}x \ll 1$),
  I find it easiest to
  leave formulas in terms of $\ln y_{\rm cut}$.
}
\begin {equation}
   \delta \left[ \frac{d\Gamma}{dx} \right] =
   \frac{\CA \alphas}{4\pi}
   \left[ \frac{d\Gamma}{dx} \right]_{\rm LO}
   \Big[
      \tfrac12 \ln^2 y_{\rm cut} + \bar s(x) \ln y_{\rm cut}
   \Bigr] .
\label {eq:sycut}
\end {equation}
The IR cut-off $y_{\rm cut}$ on $y$ is equivalent to a cut-off
$(\omega_y)_{\rm cut} = y_{\rm cut} E$ on the soft gluon energy
$\omega_y$.  In the application to fig.\ \ref{fig:LMWregion},
this cut-off would be chosen as the left boundary of the red
region, in order to get the contribution to large double and single
logs from the entire red region.  However, when comparing
results to the approach in this paper, it will be
easier to just focus on the un-integrated version, given by
the integrand of (\ref{eq:syint}).  I will loosely refer to
$\bar s(x)$ as the ``single-log coefficient,'' but this is really
short-hand for the relative coefficient of the term generating
the single log in (\ref{eq:syint}) compared to the one generating
the double log.

I should emphasize that the terms single and double log in this
paper refer solely to the dependence on the soft gluon $y$, and they
do not directly refer to whether or not particular terms
have logarithmic dependence on the underlying hard gluon energy
fraction $x$.  All the terms in (\ref{eq:sbar})
will be considered part of the ``single-log coefficient.''


\subsection {LMW logarithms for $p_\perp$ broadening}

The LMW \cite{LMW} soft radiative correction to the $\hat q$ for
$p_\perp$ broadening,
coming from the analog of the red region of fig.\ \ref{fig:LMWregion},
is%
\footnote{
  Eq.\ (\ref{eq:dqLMW}) comes from adding LMW \cite{LMW}
  eqs.\ (29) and (34) for what they call their
  ($b$) and ($a$) boundaries of the double log region.
  Then divide both sides by $\omega$,
  and integrate over $\omega$.  For more details and a notation translation
  table, see my appendix \ref{app:fix}, taking real parts throughout the
  discussion there.
  Of particular importance: the discussion in LMW's main text is in the
  context of the large-$\Nc$ approximation,
  where their quark $\hat q$ corresponds to my $\qhatA/2$.
}
\begin {subequations}
\label {eq:LMW}
\begin {equation}
  \delta \hat q_{\LMW}
  \simeq
  \hat q
  \left[
     - \frac{\CA\alphas}{2\pi} \int_{y\ll x} dy \>
         \frac{ \ln y + \bar s_{\LMW}(\Delta b) }{y}
  \right]
\label {eq:dqLMW}
\end {equation}
in my notation, with
\begin {equation}
  \bar s_{\LMW}(\Delta b) =
  2\ln\Bigl( \tfrac14 (\Delta b)^2 \sqrt{\tfrac12 \qhatA E} \Bigr)
  + 2\gammaE .
\label {eq:sbarLMW}
\end {equation}
\end {subequations}
As I now briefly review, the
transverse separation
$\Delta b$ appearing
above arises in discussions of
$p_\perp$ broadening from a slightly formal procedure.
Later, in the case of soft corrections to hard splitting
rates, it will play a more direct role.


\subsubsection {Review: the ``potential'' $V(\Delta b)$}

One way to describe the physics of fig.\ \ref{fig:LMWqhat}a is to say
that random kicks from the medium cause the $\p_\perp$ of the high-energy
particle to make a random walk in $\p_\perp$-space.  This is like
a $\p_\perp$ version of Brownian motion, and so can alternatively be
described by a $\p_\perp$ version of the diffusion equation:
\begin {equation}
   \partial_t \rho(\p_\perp,t) = \kappa \nabla_{\p_\perp}^2 \rho(\p_\perp,t) ,
\label {eq:diff}
\end {equation}
where $\rho$ is the probability distribution in $\p_\perp$.
The coefficient
$\kappa$ is the $\p_\perp$-space diffusion constant, which is related
to $\hat q$ by $\kappa = \hat q/4$.
Fourier transforming (\ref{eq:diff}) from $\p_\perp$-space to transverse
position space $\b$ gives
\begin {equation}
   \partial_t \rho(\b,t) = - C(b) \, \rho(\b,t)
\label {eq:diffb}
\end {equation}
with
\begin {equation}
  C(b) = \kappa b^2 = \tfrac14 \hat q b^2 .
\label {eq:Cb}
\end {equation}
This is the diffusive (i.e.\ $\hat q$) approximation to what
is often called, more generally, the collision kernel $C(b)$ in the
literature.%
\footnote{
  If one starts with more general considerations of elastic scattering from
  the medium, instead of the diffusion approximation (\ref{eq:diff}), one
  could start with the Fokker-Plank equation
  \[
    \partial_t \rho(\p_\perp,t) =
    \int d^2q_\perp \> \frac{d\Gamma_{\rm el}}{d^2 q_\perp} \,
       \bigl[ \rho(\p_\perp{-}\q_\perp,t) - \rho(\p_\perp,t) \bigr]
    ,
  \]
  where $\Gamma_{\rm el}$ is the cross-section for elastic scattering
  from the medium.
  By again switching to $\b$ space, this would lead to
  (\ref{eq:Cb}) with
  \[
    C(b) =
    \int d^2q_\perp \>
      \frac{d\Gamma_{\rm el}}{d^2 q_\perp} \, (1-e^{i\b\cdot\q_\perp}) .
  \]
  Formally, (\ref{eq:Cb}) is the small-$b$ approximation to this
  result.  As energy increases, deflections of the high-energy
  particles become smaller,
  and so changes in their transverse position become
  small.  The high-energy limit corresponds to the
  small-$\b$ limit and so to the $\hat q$ approximation
  (with caveats about log dependence of $\hat q$).
}

It will be useful for later discussion to multiply both sides by $i$ so
that the equation takes the mathematical form of a 2-dimensional
``Schr\"odinger equation'' (with no kinetic term):
\begin {equation}
   i \partial_t \rho(\b,t) =  V(b) \, \rho(\b,t) .
\label {eq:Schrob}
\end {equation}
with
\begin {equation}
   V(b) = - i \kappa b^2 = - \tfrac{i}{4} \hat q b^2 .
\label {eq:Vb}
\end {equation}
For this reason, the $\hat q$ approximation is sometimes called
the ``harmonic oscillator approximation.''  Note, however, that
the spring constant of this harmonic oscillator is imaginary.
Formally, (\ref{eq:Vb}) would mean that
\begin {equation}
  \hat q = 2 i V''(0) .
\label {eq:qhatfromV}
\end {equation}

\begin {figure}[t]
\begin {center}
  \includegraphics[scale=0.6]{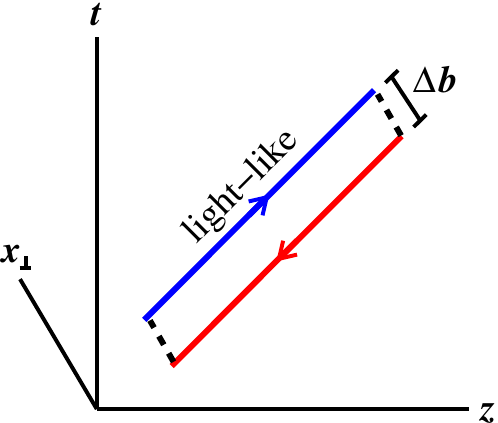}
  \caption{
     \label {fig:Wilson}
     A Wilson loop with transverse separation $\Delta b$ and
     long, light-like sides.
  }
\end {center}
\end {figure}

Analogous to how actual potential energies between static
test charges may be computed using Wilson loops, the ``potential''
$V(b)$ may be related to the type of Wilson loop shown in
fig.\ \ref{fig:Wilson} \cite{LRW1,LRW2},
which has long light-like sides, transverse
extent $\Delta b$, and
expectation $\sim e^{-i V(\Delta b) \mathbb{T}}$
(where $\mathbb{T}$ is the long time duration of the loop).
The color coding I have used for the long sides of the Wilson loop
follows the same convention as fig.\ \ref{fig:split}.
In this case, the blue line can be thought of as roughly
representing an
amplitude for a high-energy particle traveling through and interacting with the
medium, like in fig.\ \ref{fig:LMWqhat}a.  The red line can be roughly
thought of as a contribution to the conjugate amplitude, and together they
give information related to rates.  For example, formally, $\hat q$,
which contains one piece of information about rates,
can be extracted from (\ref{eq:qhatfromV}).

There are many subtle details to contend with if one wants to make
the Wilson loop definition precise.
One consideration is how exactly to implement
the different time-ordering prescriptions of the blue and red lines
while keeping the Wilson loop gauge invariant \cite{Benzke},
but that's a detail
that we'll be able to ignore.
Another, related consideration is whether one should think of
the light-like Wilson loop of fig.\ \ref{fig:Wilson} as a limiting
case of slightly {\it time}-like Wilson loops or of slightly {\it space}-like
Wilson loops.  Though one may mostly read this paper without worrying
about this detail, a little discussion here may clarify some
later arguments.
Physically, the particles represented by the
long sides of the Wilson loop
are moving slightly slower than the speed of light
(at the very least because of medium-induced masses).
On the other hand, as explained by Caron-Huot in ref.\ \cite{simon},
a very slightly {\it space}-like loop is also sensitive to
the contributions to
$\hat q$ [or more generally $C(b)$] that are due to scattering of
the high-energy particles from pre-existing fluctuations in the gauge
fields of the medium.  That includes, for example, all the physics of
elastic scattering%
\footnote{
  For the purpose of this discussion, the term ``elastic'' means that the
  high-energy particle does not split;
  it need not
  assume anything about what happens to particles in the medium.
}
from the medium.
For the slightly space-like loop, lack of
causal connection implies
that operators at different points on the loop commute with each
other, and so it matters not how the lines of the loop are time
ordered.  This will be an advantage later on for making certain
generic arguments.  (It is also the definition
relevant to a recent method for extracting $C(b)$ from lattice simulations
for weakly-coupled quark-gluon plasmas \cite{nonpertCb1,nonpertCb2}.) 
However, when it comes to calculating the soft
radiation effect $\delta\hat q$, we must instead think of the slightly time-like
loop.  And we will see later that how the lines are
time-ordered impacts the values of
different $\delta \hat q$'s arising in later application of
soft corrections to energy loss.

Overall, I will consider the bare $\hat q$ to be defined by the
slightly space-like loop (so that time-ordering does not matter), with
the additional caveat that (at the very least) no radiation effects
corresponding to $\omega_y > y_{\rm cut} E$ will be included in the
bare $\hat q$, since $y_{\rm cut} E < \omega_y \ll \omega_x$ is going
to represent the soft radition (with IR cut-off) effects that will
define $\delta\hat q$.  Fortunately, a more rigorous definition of
bare $\hat q$, and its separation from $\delta\hat q$,
will not be necessary for my purpose in this paper.


\subsubsection {LMW's calculation}

\begin {figure}[t]
\begin {center}
  \includegraphics[scale=0.35]{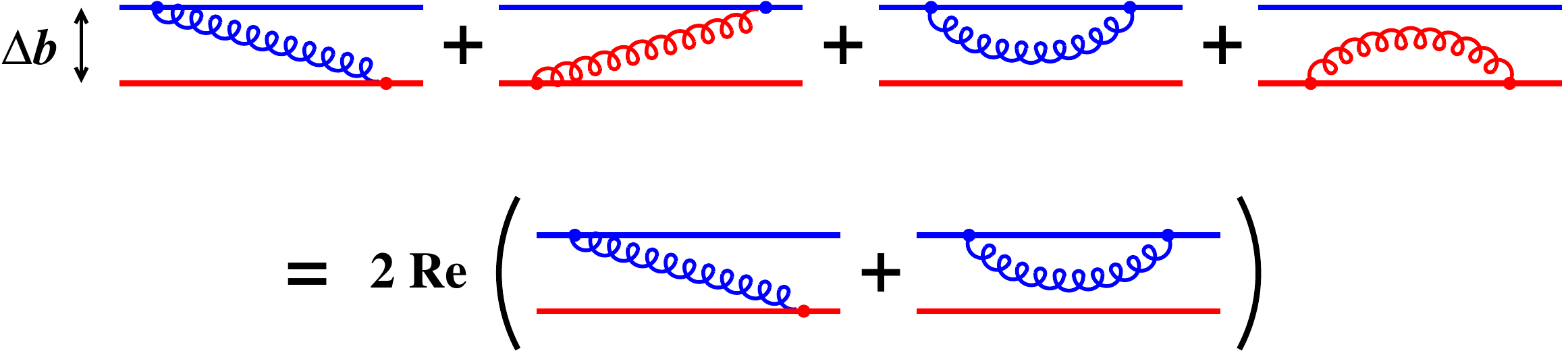}
  \caption{
     \label {fig:LMWdiags}
     The diagrams underlying the LMW calculation.  The Wilson lines and
     the radiated soft gluon line are all implicitly interacting with the
     medium.
  }
\end {center}
\end {figure}

Imagine that $\hat q$ is initially determined from scattering of the
hard particles from the medium, as in fig.\ \ref{fig:LMWqhat}a.
LMW's calculation of the soft radiative corrections of fig.\ \ref{fig:LMWqhat}b
was equivalent to computing the diagrams shown in fig.\ \ref{fig:LMWdiags}.  
In this language, their result was that
the soft-radiation correction to the initial potential (\ref{eq:Vb}) is
\begin {equation}
   \delta V_{\LMW}(\Delta b) =
   -\frac{i}{4} \delta\hat q(\Delta b) \times (\Delta b)^2 ,
\end {equation}
with $\delta\hat q(\Delta b)$ given by (\ref{eq:dqLMW}).
Since that $\delta\hat q(\Delta b)$ depends logarithmically
on $\Delta b$, it does not have a finite
limit as $\Delta b \to 0$.  That is, $\delta V_{\rm LMW}$ is
proportional to
$\bigl[\ln(\Delta b) + {\rm const}\bigr](\Delta b)^2$ at small
$\Delta b$,
not a truly quadratic potential.
In any application,
the value of $\delta\hat q$ will depend logarithmically on the relevant
scale of $\Delta b$ as so on the relevant scale of
$\Delta b$'s Fourier conjugate,
$q_\perp$.

This situation has long been known to arise even in the context of
leading-order perturbative calculations of the {\it original}
elastic-scattering $\hat q$ for a weakly-coupled quark-gluon plasma
(when running of the coupling is not included).
In that case,
it is an issue related to large-$q_\perp$ tails of the differential
Rutherford scattering cross-section and to the fact that, mathematically,
asking for the formal expectation $\langle p_\perp^2 \rangle$
(which can be dominated by very rare events with
very large $p_\perp$) can be a very
different question than asking for the ``typical''
or median value of $p_\perp^2$.%
\footnote{
   See section 3.1 of BDMPS ref. \cite{BDMPS3}.
   See also, for example, the discussion in section 1.B and
   appendix A of ref.\ \cite{HO}.
}
One may resolve these issues in the context of $p_\perp$
broadening, as briefly mentioned by LMW.  We will not need
to think about this at all, however.  For the application of this
paper, I will show that
the LMW result (\ref{eq:LMW}) for
$\delta\hat q(\Delta b)$ can be used as is, without any ambiguity of
interpretation.

\subsection {A seeming disconnect}

\begin {figure}[t]
\begin {center}
  \begin{picture}(170,70)(0,0)
    \put(3,5){\includegraphics[scale=1.0]{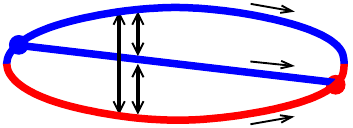}}
    \put(45,26){$b_{31}$}
    \put(73,20){$b_{23}$}
    \put(73,47){$b_{12}$}
    \put(111,65){$x_1$}
    \put(111,37){$x_2$}
    \put(111,2){$x_3$}
  \end{picture}
  \caption{
     \label {fig:LOsplitb}
     The hard splitting rate diagram,
     with emphasis on the three different transverse
     separations $b_{ij}$.  The longitudinal momentum fraction
     labels $(x_1,x_2,x_3)$ are shown for later reference.
  }
\end {center}
\end {figure}

The problem with immediately making a connection between
the LMW single log coefficient (\ref{eq:sbarLMW}) for soft radiative
corrections to $p_\perp$
broadening and the single log coefficient (\ref{eq:sbar}) extracted
from soft corrections to hard splitting processes is that the LMW
coefficient depends on one separation $\Delta b$.
For the hard splitting process, there are {\it three} different,
relevant transverse separations
\begin {equation}
   \b_{ij} \equiv \b_i - \b_j
\end {equation}
as depicted in fig.\ \ref{fig:LOsplitb}.  Moreover, these separations
are not fixed: they are functions of time.
Quantum mechanically, one must sum over all paths
the three particles can take during the emission process.
In order to see how to make use of the LMW result to include
soft corrections, we need to first drill down and review some details of
the usual leading-order
BDMPS-Z calculation of the underlying hard splitting rate.

\section {Review:
  Leading-order BDMPS-Z splitting rates in $\hat q$ approximation}
\label {sec:BDMPSZ}

\subsection {The ``Hamiltonian''}

I'll use Zakharov's version of the BDMPS-Z formalism for splitting
rates.  This corresponds to thinking of the three lines in
fig.\ \ref{fig:split} (two particles in the amplitude and one in
the conjugate amplitude) as a total of three particles evolving
forward in time.  Zakharov then treats the evolution as,
formally, a type of quantum mechanics problem.  A quick, heuristic way to
understand the basic formulation is to first ignore interactions
with the medium.  In that case, the two particles in the amplitude
will evolve quantum mechanically as $e^{-i\varepsilon_1 t}$ and
$e^{-i\varepsilon_2 t}$, where, for high-energy particles moving
nearly collinear with the $z$ axis,%
\footnote{
  For simplicity, I am assuming that the energies are high enough that
  bare and medium-induced particle masses $m_{\rm eff}$
  are ignorable compared to typical
  $p_\perp$ values for the splitting process.
  In an infinite medium, this is parametrically
  $(\omega_x \hat q)^{1/2} \gg m_{\rm eff}^2$.
}
\begin {equation}
   \varepsilon_i \simeq \sqrt{p_z^2 + p_\perp^2}
   \simeq |p_z| + \frac{p_\perp^2}{2|p_z|} \,.
\end {equation}
The one particle in the conjugate amplitude evolves instead as
$(e^{-i\varepsilon_3 t})^* = e^{+i\varepsilon_3 t}$.  Altogether, the
free system evolves as $e^{-iHt}$ with
$H = \frac{p_{\perp1}^2\vphantom{\big|}}{2|p_{z_1}|}
     + \frac{p_{\perp2}^2\vphantom{\big|}}{2|p_{z_2}|}
     - \frac{p_{\perp3}^2\vphantom{\big|}}{2|p_{z_3}|}.$
Now include (medium-averaged) interactions with the medium by
including a three-body ``potential'' analogous to the two-body
potential $V(\Delta b)$ discussed previously:
\begin {equation}
  H = \frac{p_{\perp1}^2}{2|p_{z_1}|} + \frac{p_{\perp2}^2}{2|p_{z_2}|}
     - \frac{p_{\perp3}^2}{2|p_{z_3}|}
     + V_{(3)}(\b_1,\b_2,\b_3) ,
\label {eq:H0}
\end {equation}
where $\b_i$ are the 2-dimensional transverse positions of the particles
conjugate to $\p_{\perp i}$.  There are various situations, such as
(i) the weakly-coupled limit of a quark gluon plasma or
(ii) the large-$\Nc$ limit,
where one may argue that the 3-body potential decomposes
into a sum of 2-body potentials.  However, in the context of the
harmonic oscillator (i.e.\ $\hat q$) approximation to potentials,
there is a very simple, completely general argument: Any
{\it quadratic} potential that is invariant under translations and
rotations in the transverse plane%
\footnote{
   The medium need not be invariant with respect to
   large transverse translations.  All that is relevant here
   is whether it is, to good approximation, transverse translation
   invariant over the
   scale of the tiny transverse deflections that very high-energy
   particles pick up in a formation length.
   My analysis in this paper ignores the possibility that the
   medium may not be sufficiently invariant under rotations in the
   transverse plane in situations where the jet is cutting across
   the flow of the medium.  See \cite{notransrot}.
}
can be
written in the form
\begin {equation}
   V_{(3)}(\b_1,\b_2,\b_3) = c_{12} b_{12}^2 + c_{23} b_{23}^2 + c_{31} b_{31}^2
\label {eq:V3generic}
\end {equation}
for some constants $c_{ij}$.

In this paper, I will focus on the case of $g{\to}gg$ splitting
since that's the underlying hard process for the soft radiation single
log coefficient (\ref{eq:sbar}) that I eventually want to reproduce
from LMW's corrections to momentum broadening.  In that case,
the usual decomposition used in the literature would correspond to
\begin {equation}
   V_{(3)}(\b_1,\b_2,\b_3)
   = - \tfrac{i}{8} \qhatA ( b_{12}^2 + b_{23}^2 + b_{31}^2 ) .
\label {eq:V3}
\end {equation}
The coefficients in (\ref{eq:V3}) can be motivated in various ways,
such as from arguments for weakly-coupled plasmas.
However, I will now review
a more general argument in the
context of the $\hat q$ approximation
(adapted, with some additional clarification, from
refs.\ \cite{2brem,Vqhat}%
\footnote{
  See, in particular, eq.\ (2.21) of ref.\ \cite{2brem} and the
  corresponding paragraph of appendix A of ref.\ \cite{2brem}, which cover
  the more general case where the three particles can be in any color
  representations.
}%
).

I mentioned earlier the
technical point that I was taking my bare $\hat q$ values to be
defined by the light-like limit of slightly space-like Wilson loops,
and that time-ordering prescriptions were then unimportant.
As far as $\hat q$ values are concerned, there is then no
difference between amplitude (blue) and conjugate amplitude (red)
lines in fig.\ \ref{fig:split}.  For $g{\to}gg$ splitting,
the (bare) 3-body potential in (\ref{eq:V3generic})
must then be completely symmetric under permutations, so
that
\begin {equation}
   V_{(3)} = c (b_{12}^2+b_{23}^2+b_{31}^2) .
\label {eq:V3c}
\end {equation}
(We will see later that this type of
symmetry argument does {\it not}\/ work exactly
for soft radiative corrections, where time ordering matters.)
Since the three high-energy particles
in fig.\ \ref{fig:split} must form a color singlet (after medium
averaging), the combined color representation of gluons 1 and 2
is forced to be in the adjoint representation so that that pair can
form a color singlet with gluon 3.  Now consider
the limiting case of (\ref{eq:V3c}) where $\b_1 = \b_2$.
Then the combination of gluons 1 and 2, which are on top
of each other, is indistinguishable from a single gluon at that
location.
The 3-gluon system is then equivalent to a 2-gluon system, and
so (\ref{eq:V3c}) with $\b_1 = \b_2$ must reproduce
the gluon case $-\tfrac{i}{4} \qhatA b^2$
of the 2-particle potential (\ref{eq:Vb}).
That fixes the coefficient $c$ of the 3-body harmonic oscillator
potential (\ref{eq:V3c}) to give (\ref{eq:V3}).

It will be useful to now introduce some notation that I will use
throughout the paper.  For the hard, single splitting
$E \to xE, (1{-}x)E$, I define the longitudinal momentum fractions
\begin {equation}
   (x_1,x_2,x_3) \equiv (1{-}x,x,-1) .
\label {eq:altx}
\end {equation}
$x_i$ is defined to show the flow of $p_{zi} \simeq x_i E$
forward in the time as defined by the arrows in fig.\ \ref{fig:LOsplitb}.
Note that the particle in the conjugate amplitude (red) has negative
$x_3$ in this convention.

The final step of setting up the BDMPS-Z calculation is to simplify
the 3-particle problem to an effective 1-particle problem by using
symmetries of the problem.  One may use transverse translation invariance
to eliminate one particle degree of freedom by separating out what in
ordinary 2-dimensional quantum mechanics would be the
``center of mass'' motion.  It turns out that
one may also use invariance of the original
problem under tiny rotations that change the direction of the $z$ axis
to eliminate a second particle degree of freedom.  The result is
that $\b_{12}$, $\b_{23}$ and $\b_{31}$ may be expressed in terms
of a single 2-dimensional degree of freedom $\B$, with%
\footnote{
   For a detailed discussion of this reduction to a single degree
   of freedom, in the language used
   here, see sections 2.5 and 3 of ref.\ \cite{2brem}.
   [Warning: the definition of $(x_1,x_2,x_3)$ in ref.\ \cite{2brem}
   is permuted compared to the one used here.]
   The original use of the reduction was by Zakharov \cite{Zakharov2}
   and then incorporated into BDMPS \cite{BDMS}.
   Something
   equivalent was also used by
   ref.\ \cite{AMYglue}.
   (For translations of the notation of these works, see the appendix
   of ref.\ \cite{simple}.)
}
\begin {equation}
  \b_{12} = (x_1{+}x_2) \B = -x_3 \B ,
  \quad
  \b_{23} = (x_2{+}x_3) \B = -x_1 \B ,
  \quad
  \b_{31} = (x_3{+}x_1) \B = -x_2 \B .
\label {eq:b123B}
\end {equation}
The momentum conjugate to $\B$ is
\begin {equation}
  \P = x_2 \p_{\perp 1} - x_1 \p_{\perp2}
  = x_3 \p_{\perp 2} - x_2 \p_{\perp3}
  = x_1 \p_{\perp 3} - x_3 \p_{\perp1} .
\end {equation}
With this reduction, the ``Hamiltonian'' given by (\ref{eq:H0}) and
(\ref{eq:V3}) reduces to a single, 2-dimensional harmonic oscillator
\begin {equation}
  H = \frac{P^2}{2M_0} + \tfrac12 M_0 \Omega_0^2 B^2
\label {eq:H}
\end {equation}
with
\begin {equation}
  M_0 = |x_1 x_2 x_3| E = x(1{-}x)E ,
\end {equation}
and (for $g{\to}gg$)
\begin {equation}
  \Omega_0 =
  \sqrt{ - \frac{i\qhatA}{2 E}
         \Bigl( \frac{1}{x_1} + \frac{1}{x_2} + \frac{1}{x_3} \Bigr) }
  =
  \sqrt{ - \frac{ i\qhatA(1{-}x{+}x^2) }{ x(1{-}x)E } } .
\label {eq:Om0}
\end {equation}
Note that $M_0$ and $\Omega_0$ are both symmetric under permutation
of the three momentum fractions $(x_1,x_2,x_3)$.


\subsection {The calculation}

\begin {figure}[t]
\begin {center}
  \begin{picture}(220,70)(0,0)
    \put(5,27){$\displaystyle{
        \frac{d\Gamma}{dx} = 2\Re \int_0^\infty \!\! d(\Delta t)
    }$}
    \put(115,5){\includegraphics[scale=0.6]{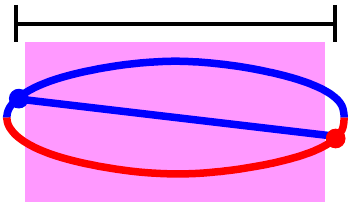}}
    \put(160,60){$\Delta t$}
  \end{picture}
  \caption{
     \label {fig:LOrate}
     The hard splitting rate.  The shaded region indicates the
     region of time evolution between the two vertices, which is
     described by the 2-dimensional ``quantum mechanics'' Hamiltonian $H$.
  }
\end {center}
\end {figure}
Following Zakharov's formulation \cite{Zakharov2}, the rate is
given schematically in fig.\ \ref{fig:LOrate} which, in my notation,
translates to
\begin {equation}
   \left[ \frac{d\Gamma}{dx} \right]_{\rm LO} =
   \frac{\alphas P(x)}{M_0^2} \Re \int_0^\infty \!\! d(\Delta t) \>
      \grad_\B \cdot \grad_{\B'}
      \langle \B,\Delta t|\B',0\rangle \Bigl|_{\B=0=\B'} .
\label {eq:Zrate}
\end {equation}
Here, the subscript ``LO'' 
stands for leading order in powers of ``bremsstrahlung $\alphas$.''
The factor $\langle \B,\Delta | \B',0 \rangle$ corresponds to time
evolution in the shaded region of fig.\ \ref{fig:LOrate} and is
given by the quantum mechanics propagator associated with the
Hamiltonian (\ref{eq:H}).  In the $\hat q$ approximation I am using in
this paper, this is simply a standard harmonic oscillator propagator,
which in two dimensions is
\begin {equation}
  \langle \B,\Delta t|\B',0\rangle
  = \frac{M_0 \Omega_0 \csc(\Omega_0 \Delta t)}{2\pi i} \,
    \exp\Bigl(
       \tfrac{i}{2} M_0 \Omega_0
       \bigl[
          (B^2{+}B'^2) \cot(\Omega_0 \Delta t)
          - 2\B\cdot\B' \csc(\Omega_0\Delta t)
       \bigr]
    \Bigr) .
\label {eq:HOprop}
\end {equation}
The derivatives, the overall factor of $\alphas$, and the DGLAP splitting
function $P(x)$ in (\ref{eq:Zrate})
come from the two high-energy splitting vertices in the
diagram.  The current that a transversely-polarized gluon couples to
in the collinear limit relevant to high energies is proportional
to the transverse momentum.
Correspondingly, in momentum space, the derivatives in (\ref{eq:Zrate})
correspond to factors of $\P$, which characterize transverse momentum
and which become $-i\grad$ in $\B$ space.
The factor of $1/M_0^2$ in (\ref{eq:Zrate}) arises from various
normalization factors.

Using (\ref{eq:HOprop}) in (\ref{eq:Zrate}) and integrating over $\Delta t$
then yields
\begin {equation}
   \left[ \frac{d\Gamma}{dx} \right]_{\rm LO} =
   \frac{\alphas P(x)}{\pi} \Re(i\Omega_0) .
\label {eq:LOrate}
\end {equation}
For the case of $g{\to}gg$, (\ref{eq:Om0}) then gives the final,
standard result
\begin {equation}
   \left[ \frac{d\Gamma}{dx} \right]_{\rm LO} =
   \frac{\alphas P_{g\to gg}(x)}{2\pi}
        \sqrt{ \frac{\qhatA(1{-}x{+}x^2)}{x(1{-}x)E} }
\end {equation}
quoted earlier for the $\hat q$ approximation in an infinite medium.


\section{Incorporating LMW $\hat q$ corrections into the BDMPS-Z calculation}
\label{sec:calculation}

\subsection {Setup}

The idea is to incorporate the LMW corrections to $\hat q$ into the
3-gluon potential (\ref{eq:V3}) used for the BDMPS-Z calculation,
choosing the $\Delta b$ in LMW to be the $b_{ij}$ appearing, respectively,
in each term of the potential:
\begin {equation}
   V_{(3)}^{\rm eff}(\b_1,\b_2,\b_3)
   = - \tfrac{i}{8} \Bigl[
          \qhatA^{\rm eff}(b_{12}) \, b_{12}^2
          + \qhatA^{\rm eff}(b_{23}) \, b_{23}^2
          + \qhatA^{\rm eff}(b_{31}) \, b_{31}^2
       \Bigr] .
\label {eq:V3eff}
\end {equation}
The assumption here that the $\qhatA^{\rm eff}$ in the first term cares
only about the separation $b_{12}$ can be justified in the large-$\Nc$ limit.

\begin {figure}[t]
\begin {center}
  \includegraphics[scale=0.4]{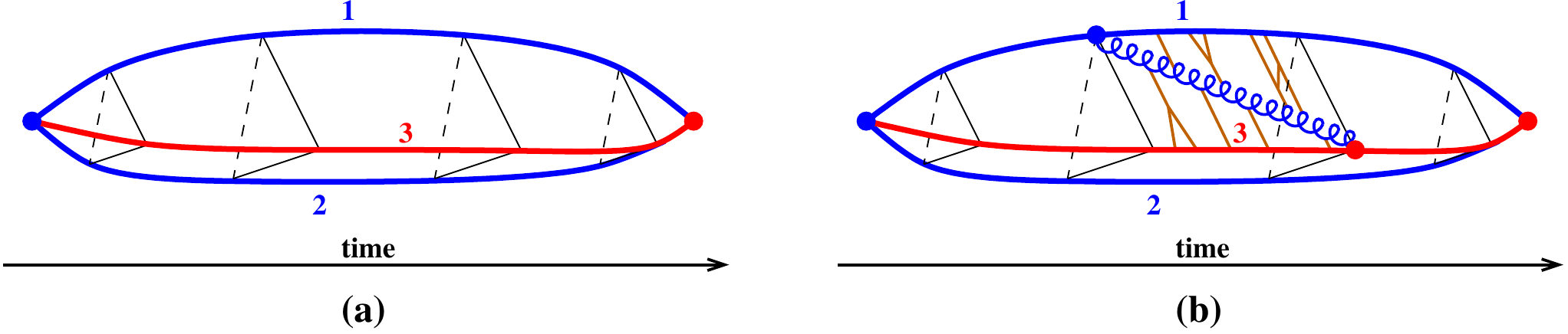}
  \caption{
     \label {fig:lozenge}
     (a) A redrawing of the time-ordered
     $g{\to}gg$ hard splitting diagram of
     fig.\ \ref{fig:split}b, depicted here as a 3-dimensional
     lozenge with triangular
     cross-section.  In the large-$\Nc$ limit, additions to this diagram
     representing interactions must be drawable on the surface of the
     lozenge.  The black triangles are just a visual aid, representing
     cross-sections of the lozenge at instants in time.
     (b) An example of a soft correction (curly gluon line), along with
     an example of medium correlations represented by the brown lines.
     Both 2-point and higher-point correlations are shown by way of
     example, but the analysis in this paper only assumes that
     the correlation length of the medium is small compared to formation
     times, not that the medium is weakly-coupled.  Drawing
     the soft correction and correlators on the
     surface of the lozenge is just an abstraction for the sake of
     discussing the dominant time-ordered diagrammatic
     contributions in the large-$\Nc$ limit: there is
     {\it no}\/ implication that the physical path of these lines
     must follow such a surface in transverse position space $\b$.
  }
\end {center}
\end {figure}

To see this, imagine re-drawing the original $g{\to}gg$ time-ordered
rate diagram of fig.\ \ref{fig:split}b as a triangular cross-section
lozenge, as in fig.\ \ref{fig:lozenge}a.
The large-$\Nc$ requirement that diagrams be planar can be understood
as a requirement that any additions to fig.\ \ref{fig:lozenge}a must
lie on the surface of the lozenge without crossing lines.
(The surface of the lozenge is
topologically equivalent to a 2-sphere, which can be stereographically
projected onto a plane.  So any diagram that can be drawn on the
lozenge's surface without crossing lines can be mapped to a planar diagram.)
Now consider, in particular, a soft correction to the hard splitting,
as in fig.\ \ref{fig:LOsplit+SOFT}.
Fig.\ \ref{fig:lozenge}b gives an example, where the soft curly gluon line
connects lines 1 and 3.  In the large-$\Nc$ limit, the soft line must then
lie along the corresponding face of the lozenge.
Correlations of interactions with the medium may be represented by
a network of medium gluon correlators connecting
to the high-energy particles.  In large-$\Nc$, these correlations
(brown lines in the figure) must also lie on the surface of
the lozenge.  That means that the soft gluon in
fig.\ \ref{fig:lozenge}b only has medium correlations with particles
1 and 3 in this example.  That soft gluon line is not affected at
all by particle 2 and so only knows about the separation $\b_{31}$
between particles 1 and 3.  Fig.\ \ref{fig:lozenge}b represents
a correction to {\it direct}\/ medium correlations between particles
1 and 3 and so represents a correction to the $\qhatA b_{13}^2$ term
of the original potential (\ref{eq:V3}).  In summary, the first term
in the corrected potential (\ref{eq:V3eff}) depends only on $b_{31}$ in the
large-$\Nc$ limit.

We now need to repeat the
BDMPS-Z calculation using the effective
potential (\ref{eq:V3eff}) that includes LMW soft-radiative corrections
to $\hat q$.  This is no longer a harmonic oscillator problem, but we
can still get a relatively simple answer to first order in the
high-energy splitting $\alphas$ by expanding around the usual
BDMPS-Z $\hat q$-approximation result using time-ordered quantum
mechanical perturbation theory in the correction $\delta V$
to the original potential $V_{(3)}$:
\begin {equation}
   \delta V_{(3)}(\b_1,\b_2,\b_3)
   = - \tfrac{i}{8} \Bigl[
          \delta \qhatA^{\scriptscriptstyle\LMW}\!(b_{12}) \, b_{12}^2
          + \delta\qhatA^{\LMW}\!(b_{23}) \, b_{23}^2
          + \delta\qhatA^{\LMW}\!(b_{31}) \, b_{31}^2
       \Bigr] .
\end {equation}
Using the reduction (\ref{eq:b123B}) of the 3-particle problem to an
effective 1-particle problem, that's
\begin {equation}
   \delta V_{(3)}(B)
   = - \tfrac{i}{8} \Bigl[
          \delta \qhatA^{\LMW}\!(|x_3|B) \, (x_3 B)^2
          + \delta\qhatA^{\LMW}\!(|x_1|B) \, (x_1 B)^2
          + \delta\qhatA^{\LMW}\!(|x_2|B) \, (x_2 B)^2
       \Bigr] .
\label {eq:dV3eff}
\end {equation}

\begin {figure}[t]
\begin {center}
  \begin{picture}(340,100)(0,0)
    \put(5,66){$\displaystyle{ \delta }$}
    \put(20,44.5){\includegraphics[scale=0.6]{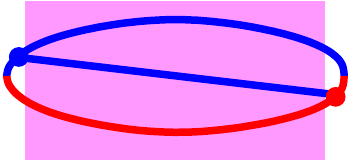}}
    \put(130,66){$\displaystyle{
        = -i \int_0^{\Delta t} \!\! dt_1 \int d^2B_1
    }$}
    \put(240,35){\includegraphics[scale=0.6]{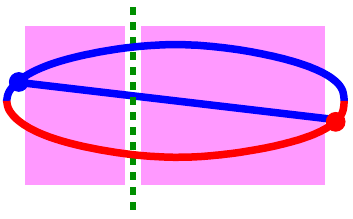}}
    \put(267,25){$\Delta V_{(3)}$}
    \put(260,13){insertion}
    \put(257,3){at $(t_1,\B_1)$}
  \end{picture}
  \caption{
     \label {fig:dZrate}
     First-order perturbative correction to the integrand of
     the leading-order
     hard splitting rate in fig.\ \ref{fig:LOrate}.
  }
\end {center}
\end {figure}

First order perturbation theory for (\ref{eq:Zrate}) corresponds to%
\footnote{
  This is similar in form to the method used by Mehtar-Tani and Tywoniuk
  \cite{IOE2}
  to deal with logarithmic dependence (the Rutherford tail)
  in the {\it bare} value of
  $\hat q$ through what they call the Improved Optical Expansion.
  Here, however, I am taking the bare $\hat q$ to be fixed and am
  instead interested in the LMW soft radiative corrections, which
  have logarithmic dependence, and my expansion parameter is the
  assumed-small $\alphas$ associated with the splitting of high-energy
  particles (including what I call the ``soft'' ones).
}
\begin {multline}
   \delta \left[ \frac{d\Gamma}{dx} \right] =
   \frac{\alphas P(x)}{M_0^2} \Re \int_0^\infty \!\! d(\Delta t)
      \int_0^{\Delta t} dt_1 \int d^2B_1 \>
\\ \times
      \grad_\B \cdot \grad_{\B'}
      \Bigl[
         \underbrace{ \langle \B,\Delta t | \B_1,t_1 \rangle }_{\text{HO osc.}}
         \bigl( -i \, \delta V_{(3)}(B_1) \bigr)
         \underbrace{ \langle \B_1,t_1 | \B',0 \rangle }_{\text{HO osc.}}
      \Bigr]
    \Big|_{\B=0=\B'} \,,
\label {eq:dZrate}
\end {multline}
which I've depicted schematically in fig.\ \ref{fig:dZrate}.
It's convenient to switch integration variables to
$(t,t') \equiv (\Delta t{-}t_1, -t_1)$ and then reorganize (\ref{eq:dZrate})
as
\begin {multline}
   \delta\left[ \frac{d\Gamma}{dx} \right] =
   \frac{\alphas P(x)}{M_0^2}
   \Re \biggl\{
      -i \int d^2B_1 
      \left[
         \int_0^\infty dt \> \grad_\B \langle \B,t | \B_1,0 \rangle
      \right]_{\B=0}
\\
      \cdot
      \left[
         \int_{-\infty}^0 dt' \> \grad_{\B'} \langle \B_1,0 | \B',t' \rangle
      \right]_{\B'=0}
      \delta V_{(3)}(B_1)
   \biggr\} . 
\label {eq:dZrate2}
\end {multline}
Using the harmonic oscillator propagator (\ref{eq:HOprop}), the time integrals
give%
\footnote{
  These are the same integrals that appear in section 5.1 of ref.\ \cite{2brem}
  for the analysis of the initial and final 3-particle evolution in the
  context of {\it hard}\/ radiative corrections to single splitting.
}
\begin {equation}
   \left[
      \int_0^\infty dt \> \grad_\B \langle \B,t | \B_1,0 \rangle
   \right]_{\B=0}
   = - \frac{i M_0 \B_1}{\pi B_1^2} e^{-\frac12 M_0 \Omega_0 B_1^2}
   =
   \left[
      \int_{-\infty}^0 dt_1 \> \grad_{\B'} \langle \B_1,0 | \B',t' \rangle
   \right]_{\B'=0}
   ,
\end {equation}
and so (\ref{eq:dZrate2}) simplifies to
\begin {equation}
   \delta\left[ \frac{d\Gamma}{dx} \right] =
   \frac{\alphas P(x)}{\pi^2}
   \Re \biggl\{
      i \int \frac{d^2B_1}{B_1^2} \> e^{-M_0\Omega_0 B_1^2}
      \, \delta V_{(3)}(B_1)
   \biggr\} . 
\label {eq:dZrate3}
\end {equation}

\subsection {First (naive) calculation}

Now use (\ref{eq:dZrate3}) with the $\delta V_{(3)}$ of
(\ref{eq:dV3eff}) and the LMW soft radiative correction
(\ref{eq:LMW}) to $\hat q$.
For the moment, I will use (\ref{eq:LMW}) for {\it all}\/ of
the $\delta \hat q$'s.  But we will need to revisit that choice later
for $\delta\hat q(b_{12})$, which involves two {\it blue} lines
in fig.\ \ref{fig:lozenge}a rather than
a blue line paired with a red line.

Because of the symmetric treatment of different pairs of the hard
particles in the $\delta V_{(3)}$ potential (\ref{eq:dV3eff}), it's
useful to correspondingly break up the rate correction (\ref{eq:dZrate3})
into the terms coming from each such pair:
\begin {subequations}
\label {eq:channels}
\begin {equation}
   \delta\left[ \frac{d\Gamma}{dx} \right] =
   \delta\left[ \frac{d\Gamma}{dx} \right]_{12}
   + \delta\left[ \frac{d\Gamma}{dx} \right]_{23}
   + \delta\left[ \frac{d\Gamma}{dx} \right]_{31}
\label {eq:decompose}
\end {equation}
with
\begin {equation}
   \delta\left[ \frac{d\Gamma}{dx} \right]_{ij} =
   \frac{\alphas P(x)}{8\pi^2}
   x_k^2
   \Re
      \int d^2B_1 \> e^{-M_0\Omega_0 B_1^2}
      \, \delta\qhatA^\LMW\!(|x_k| B_1) .
\end {equation}
\end {subequations}
Above and hereafter, $k$ refers to the index $1,2,3$ that is
{\it different}\/ from both
$i$ and $j$.
Applying (\ref{eq:dqLMW}), I'll write (\ref{eq:channels}) using the notation
\begin {equation}
   \delta \left[ \frac{d\Gamma}{dx} \right]_{ij} =
   \int_{y \ll x} dy \>
   \left[ \frac{d\Gamma}{dx\,dy} \right]_{ij} ,
\label {eq:dGij}
\end {equation}
with
\begin {equation}
   \left[ \frac{d\Gamma}{dx\,dy} \right]_{ij}
   =
   - \frac{\CA \alphas^2 P(x)}{16\pi^2 y}
   x_k^2 \qhatA
   \Re
      \int_0^\infty d(B_1^2) \> e^{-M_0\Omega_0 B_1^2}
      \bigl[ \ln y + \bar s_\LMW(|x_k| B_1) \bigr]
\label {eq:dZrateij}
\end {equation}
in the small-$y$ limit relevant to the soft radiative corrections
(\ref{eq:dGij}).  (The notation $d\Gamma/dx\,dy$ also makes contact
with the notation of refs.\ \cite{qcd,logs} for double splitting
with overlapping formation times.)
With (\ref{eq:sbarLMW}) for $\bar s_\LMW$, the $B_1^2$
integral in (\ref{eq:dZrateij}) is straightforward,%
\footnote{
  Change integration variable to $\lambda \equiv M_0\Omega_0 B_1^2$
  and use $\int_0^\infty d\lambda \> e^{-\lambda} \ln\lambda = -\gammaE$.
  Note that the $\gammaE$ term from this integration cancels the
  $\gammaE$ term in LMW's single log coefficient (\ref{eq:sbarLMW})!
}
giving
\begin {equation}
   \left[ \frac{d\Gamma}{dx\,dy} \right]_{ij} =
   - \frac{\CA \alphas^2 P(x)}{4\pi^2 y}
   \Re \left\{
      \frac{x_k^2 \qhatA}{4M_0\Omega_0}
      \left[
        \ln y +
        2\ln\left( \frac{x_k^2}{4M_0\Omega_0} \sqrt{ \tfrac12 \qhatA E } \right)
      \right]
   \right\} .
\label {eq:dGij2}
\end {equation}

This result can be algebraically manipulated into a nicer form by
using (\ref{eq:Om0}) to show that
\begin {equation}
   \frac{x_k^2 \qhatA}{4 M_0\Omega_0} = i w_{ij} \Omega_0
\label{eq:wrelation}
\end {equation}
with
\begin {equation}
   w_{ij} \equiv \frac{x_k^2}{x_1^2+x_2^2+x_3^2} \,.
\label {eq:wdef}
\end {equation}
By definition, the $w_{ij}$ have the property that
\begin {equation}
   w_{12} + w_{23} + w_{31} = 1 ,
\end {equation}
and we'll see later that it is useful to think of them as relative
weights of various contributions (hence the choice of letter ``$w$'').
For now, use (\ref{eq:wrelation}) to rewrite (\ref{eq:dGij2}) as
\begin {equation}
   \left[ \frac{d\Gamma}{dx\,dy} \right]_{ij} =
   - \frac{\CA \alphas^2 P(x)}{4\pi^2 y} \, w_{ij}
   \Re \left\{
      i\Omega_0
      \left[
        \ln y +
        2\ln\left( i w_{ij} \Omega_0 \sqrt{ E/2\qhatA } \right)
      \right]
   \right\} .
\label {eq:dGij3}
\end {equation}
Knowing the complex phase of $\Omega_0 = e^{-i\pi/4}|\Omega_0|$, this can be
rewritten as
\begin {equation}
   \left[ \frac{d\Gamma}{dx\,dy} \right]_{ij} =
   - \frac{\CA \alphas}{4\pi y} \, w_{ij}
   \left[ \frac{d\Gamma}{dx} \right]_{\rm LO}
      \left[
        \ln y +
        2\ln\left( w_{ij} |\Omega_0| \sqrt{ E/2\qhatA } \right)
        - \frac{\pi}{2}
      \right]
\label {eq:dGij4}
\end {equation}
in terms of the BDMPS-Z rate
$[ d\Gamma/dx ]_{\rm LO}$ given by (\ref{eq:LOrate}).
The $\pi/2$ term arises from the logarithm of the complex phase in
(\ref{eq:dGij3}) [in combination with the operation
$2\Re\{ i\Omega_0 \cdots \}$].
The corresponding single-log coefficient $\bar s(x)$ appearing in
(\ref{eq:syint}) for soft radiative corrections to hard $g{\to}gg$ splitting
would then be
\begin {equation}
  \bar s = \bar s_{12} + \bar s_{23} + \bar s_{31}
\label {eq:sbar123}
\end {equation}
with
\begin {equation}
  \bar s_{ij} =
  w_{ij}
  \left[
    2\ln\left( w_{ij} |\Omega_0| \sqrt{ E/2\qhatA } \right)
          - \frac{\pi}{2}
  \right] .
\label {eq:sij}
\end {equation}
Using the explicit values $(x_1,x_2,x_3) = (1{-}x,x,-1)$ of the
three longitudinal momentum fractions, and using the formula
(\ref{eq:Om0}) for $\Omega_0$, one may algebraically manipulate
this result into a form similar to the coefficient
(\ref{eq:sbar}) extracted from the soft-$y$ limit of difficult
generic-$y$ calculations in ref.\ \cite{logs}.
By having instead repeated BDMPS-Z using the LMW correction to $\hat q$,
we obtain here the slightly different result
\begin {equation}
  \bar s(x) =
      - \ln\bigl(16\,x(1{-}x)(1{-}x{+}x^2)\bigr)
      + 2 \, \frac{ \bigl[
                 x^2 \bigl( \ln x - \frac{\pi}{8} \bigr)
                 + (1{-}x)^2 \bigl( \ln(1{-}x) - \frac{\pi}{8} \bigr)
                 {\red{ {} - \frac{\pi}{8} }}
               \bigr] }
             { (1-x+x^2) }
      .
\label {eq:sbarBAD}
\end {equation}
This result matches (\ref{eq:sbar}) {\it except} for the very last
$\pi/8$ term above (in red).
This discrepancy originates from the $\pi/2$ term in (\ref{eq:sij}) for
the particular case of $\bar s_{12}$.


\subsection {Fixing up amplitude-amplitude $\delta\hat q(b_{12})$}
\label {sec:fix}

As mentioned earlier, LMW's calculation \cite{LMW} of soft radiative
corrections to $p_\perp$ broadening corresponds to studying the rate
of $p_\perp$ change for a single particle, and a rate involves
an amplitude for the particle (a blue line in my conventions) multiplied
by a conjugate amplitude for that particle (a red line).
However, in the previous derivation, in one place I
used LMW's formula for $\delta\hat q$ to treat soft radiation between two
particles in the amplitude (two blue lines).  We now need to go back
and fix that up.  Fortunately, LMW's derivation can be adapted
to this case.

\begin {figure}[t]
\begin {center}
  \includegraphics[scale=0.35]{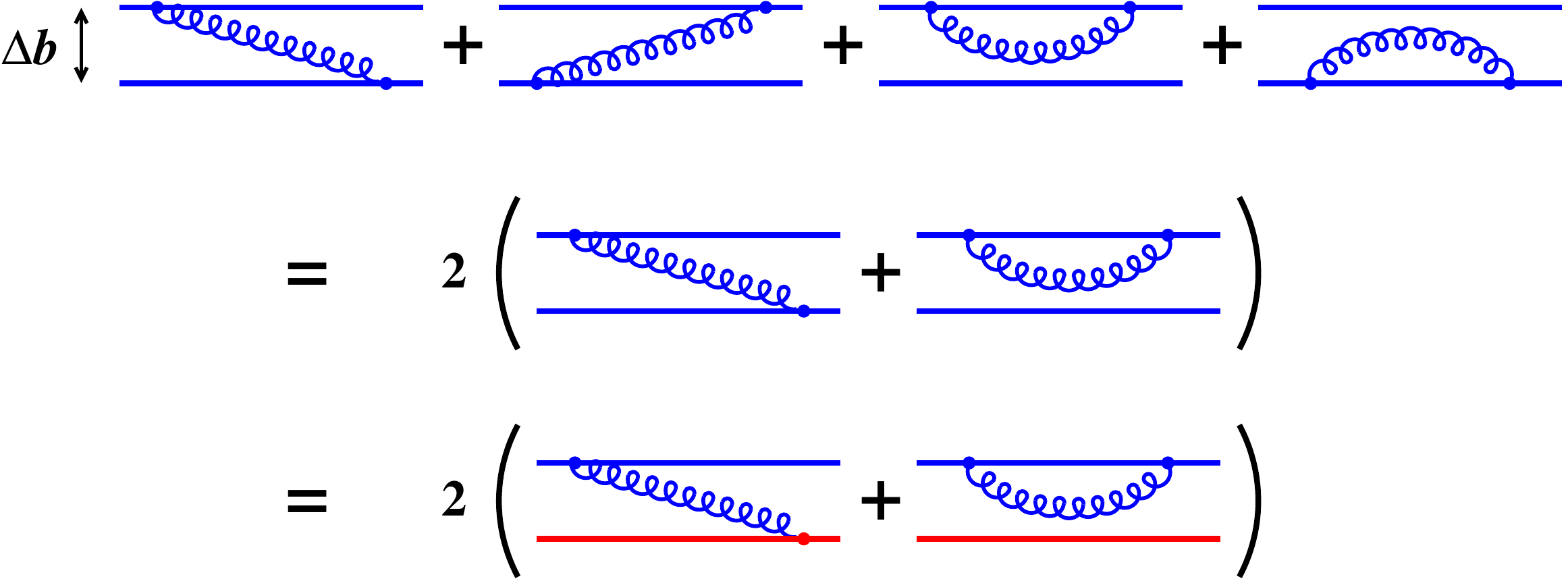}
  \caption{
     \label {fig:LMWdiags12}
     Similar to the diagrams of fig.\ \ref{fig:LMWdiags} for the LMW
     correction, but here both Wilson lines are amplitude (blue) lines.
  }
\end {center}
\end {figure}

The top line of fig.\ \ref{fig:LMWdiags12} shows the analog, for two
amplitude Wilson lines, of my depiction of LMW's diagrams in
fig.\ \ref{fig:LMWdiags}.
By rotation invariance about the direction of the Wilson lines,
the sum can be rewritten as in the second line of fig.\ \ref{fig:LMWdiags12}.
Finally, LMW's calculation is determined by the soft gluon propagator.
It matters whether that propagator is in the amplitude or
conjugate amplitude (blue or red), which it inherits in these
diagrams from the first vertex it is emitted from.  In the
last two lines of fig.\ \ref{fig:LMWdiags12}, it does not matter
whether the lower Wilson line is colored blue or red.
By comparing the last line of this figure to the LMW case of
fig.\ \ref{fig:LMWdiags}, we then see that the result is the same
{\it except}\/ that one should not take the real part $\Re(\cdots)$
at the end.  We can use the derivation from LMW's paper
\cite{LMW} if we (i) avoid ever taking
the real part and, correspondingly, (ii) are very careful to
keep track of complex phases in their derivation.
See appendix \ref{app:fix} for details.
The result is that (\ref{eq:sbarLMW}) is modified to
\begin {equation}
  \bar s_{\mbox{\scriptsize blue-blue}}(\Delta b) =
  2\ln\Bigl(
      \tfrac14 (\Delta b)^2 \sqrt{\tfrac12 \qhatA E}
      \, e^{-i\pi/4}
  \Bigr)
  + 2\gammaE .
\label {eq:sblueblue}
\end {equation}
The only difference is the factor of $e^{-i\pi/4}$ inside the argument of the
logarithm, and so
\begin {equation}
  \bar s_{\mbox{\scriptsize blue-blue}}(\Delta b) =
  \bar s_\LMW(\Delta b) - \frac{i\pi}{2}.
\label {eq:bbvsLMW}
\end {equation}
Now using this amplitude-amplitude soft correction in (\ref{eq:dZrateij})
in the case of $[d\Gamma/dx\,dy]_{12}$ gives
\begin {equation}
   \left[ \frac{d\Gamma}{dx\,dy} \right]_{12} =
   - \frac{\CA \alphas^2 P(x)}{4\pi^2 y}
   \Re \left\{
      \frac{x_3^2 \qhatA}{4M_0\Omega_0}
      \left[
        \ln y +
        2\ln\left(
          \frac{x_3^2}{4M_0\Omega_0} \sqrt{ \tfrac12 \qhatA E } \, e^{-i\pi/4}
        \right)
      \right]
   \right\}
\end {equation}
instead of (\ref{eq:dGij2}).  The explicit $e^{-i\pi/4}$ above cancels the
phase of $\Omega_0$ inside the logarithm, eliminating the $\pi$ terms
in this case, so that the corresponding version of
(\ref{eq:sij}) is 
\begin {equation}
  \bar s_{12} =
  w_{12}
  \left[
    2\ln\left( w_{12} |\Omega_0| \sqrt{ E/2\qhatA } \right)
  \right] .
\label {eq:s12}
\end {equation}

\subsection {Total}

Using (\ref{eq:sij}) for blue-red pairs and (\ref{eq:s12}) for
the blue-blue pair, the total single log coefficient is
\begin {multline}
   \bar s = \bar s_{12} + \bar s_{23} + \bar s_{31}
\\
   =
  w_{12}
  \left[
    2\ln\left( w_{12} |\Omega_0| \sqrt{ E/2\qhatA } \right)
  \right]
  +
  w_{23}
  \left[
    2\ln\left( w_{23} |\Omega_0| \sqrt{ E/2\qhatA } \right)
          - \frac{\pi}{2}
  \right]
\\
  +
  w_{31}
  \left[
    2\ln\left( w_{31} |\Omega_0| \sqrt{ E/2\qhatA } \right)
          - \frac{\pi}{2}
  \right]
  .
\label {eq:sfinal}
\end {multline}
This now exactly reproduces the result (\ref{eq:sbar}) extracted from
the small-$y$ limit of generic-$y$ results for double splitting.


\section {Other ways to write the final answer}
\label {sec:rewrites}

Algebraically,
the final answer (\ref{eq:sfinal}) may be evocatively written
directly in terms of the LMW single-log coefficient of (\ref{eq:sbarLMW})
as
\begin {subequations}
\begin {equation}
  \bar s =
  w_{12} \, \bar s_\LMW( \bar b_{12} )
  + w_{23} \bigl[ \bar s_\LMW( \bar b_{23} ) - \tfrac{\pi}{2} \bigr]
  + w_{31} \bigl[ \bar s_\LMW( \bar b_{31} ) - \tfrac{\pi}{2} \bigr] ,
\label {eq:sbbar}
\end {equation}
where
\begin {equation}
  \bar b_{ij} \equiv
  \sqrt{ w_{ij} \, \frac{4|\Omega_0|}{e^{\gammaE} \qhatA} }
\label {eq:bbarij}
\end {equation}
\end {subequations}
may be interpreted as a typical separation of the indicated pair
during the underlying hard splitting process.

For qualitative understanding of this formula, it will also be useful to
use (\ref{eq:wrelation}) and (\ref{eq:wdef}) to rewrite it in the form
\begin {subequations}
\begin {equation}
  \bar s =
  w_{12} \, \bar s_\LMW( |x_3 \bar B| )
  + w_{23} \bigl[ \bar s_\LMW( |x_1 \bar B| ) - \tfrac{\pi}{2} \bigr]
  + w_{31} \bigl[ \bar s_\LMW( |x_2 \bar B| ) - \tfrac{\pi}{2} \bigr]
\end {equation}
with
\begin {equation}
  \bar B \equiv \sqrt{ \frac{1}{e^{\gammaE} M_0 \Omega_0} } .
\label {eq:Bbar}
\end {equation}
\end {subequations}

I have no physical insight to offer about the $O(1)$
normalization factor $e^{\gammaE}$
in (\ref{eq:bbarij}) and (\ref{eq:Bbar}):
I simply chose that normalization so that
(\ref{eq:sbbar}) would reproduce (\ref{eq:sfinal}).
However, one may understand both the parametric scale
and longitudinal momentum fraction $(x_1,x_2,x_3)$ dependence
of the $\bar b_{ij} = |x_k \bar B|$.  First, consider the
$x \ll 1$ case of the hard $g{\to}gg$ splitting (but $x \gg y$ so
that $x$ is still hard compared to the soft corrections we have been
computing).  In that case, the $x_2{=}x$ gluon is the hard particle
most deflected by the medium, and its deflection is what controls
the formation time, so that
$t_{\rm form}(x) \sim \sqrt{\omega_x/\hat q} = \sqrt{xE/\hat q}$.
The transverse momentum kicks during the formation time of the
hard splitting process are then of order
$Q_\perp \sim \sqrt{ \hat q t_{\rm form}} \sim (x E \hat q)^{1/4}$,
and the corresponding transverse separation scale should
be $1/Q_\perp \sim (x E \hat q)^{-1/4}$ for the separation of the
easily-deflected $x_2{=}x$ gluon from the harder-to-deflect
$x_1$ and $x_3$ gluons.  So we expect
\begin {equation}
   b_{13} \ll b_{12} \simeq b_{23} \sim (x E \hat q)^{-1/4}
   \quad \mbox{for $x \ll 1$} ,
\label {eq:smallx}
\end {equation}
which is indeed the parametric behavior of (\ref{eq:bbarij})
in this limit.  But we can make a more precise argument about
the $x$ dependence, without assuming $x \ll 1$, by remembering
that transverse separations are precisely related to the
reduced variable $\B$ by (\ref{eq:b123B}): namely,
$b_{ij} = |x_k \B|$.  The variable $\B$, in turn, describes a
harmonic oscillator with mass $M_0$ and frequency $\Omega_0$.
For a quantum harmonic oscillator, the fundamental distance scale is
parametrically $(M_0\Omega_0)^{-1/2}$, as in
(\ref{eq:Bbar}).  So, in hindsight, we could have expected
that the typical separations $b_{ij}$ would be proportional to
the $\bar b_{ij} = |x_k \bar B|$ determined by (\ref{eq:Bbar}).

\begin {figure}[t]
\begin {center}
  \includegraphics[scale=0.4]{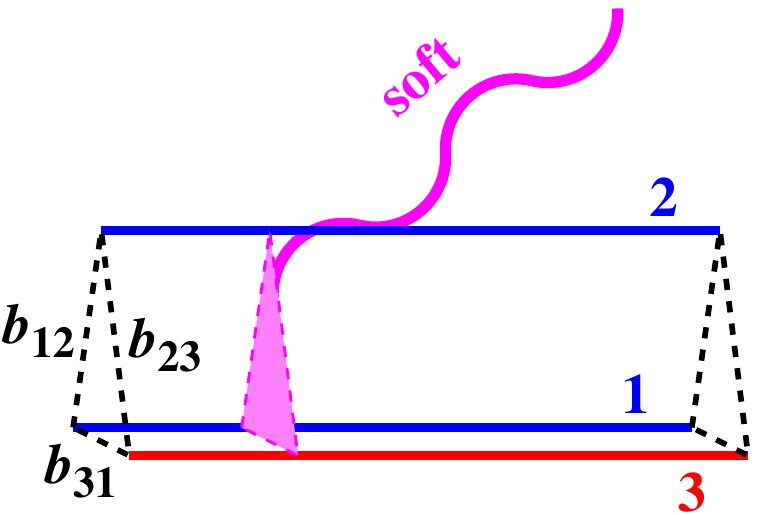}
  \caption{
     \label {fig:xsmall}
     Soft radiation from three hard lines somewhere in the middle of
     a hard $g{\to}gg$ splitting process.
     This is a magnified view of one of the soft vertices in
     fig.\ \ref{fig:LOsplit+SOFT}, and here the magenta triangle
     is just a way to represent that the soft gluon can couple
     to any one of the three hard lines. 
     This figure depicts
     the relative transverse positions of the lines during the
     soft emission process and shows the case of
     slightly small $x$, for which $x_2{=}x$ is typically
     the farthest away of the three hard lines.
  }
\end {center}
\end {figure}

I originally introduced the weights $w_{ij}$ appearing in
(\ref{eq:sbbar}) as the relative sizes (\ref{eq:wdef})
of squares of the longitudinal momentum fractions $(x_1,x_2,x_3)$.
For some insight into their role, use (\ref{eq:bbarij}) to
re-express the weights in terms of the relative sizes of
the typical squared transverse separations:
\begin {equation}
   w_{ij} =
   \frac{ \bar b_{ij}^2 }
        { \vphantom{\bigl|^*}
          \bar b_{12}^2 + \bar b_{23}^2 + \bar b_{31}^2 } \,.
\label {eq:wdefb}
\end {equation}
Now consider the picture in fig.\ \ref{fig:xsmall} of a soft emission
from the three lines of an underlying hard emission.  In this figure,
the transverse separations of the hard lines are depicted near
the time of the soft emission.
Imagine two of those lines are relatively close together, as
in the case $x \ll 1$ covered by
(\ref{eq:smallx}).  It will be harder for an even softer
(and so long wavelength) emission $y$ to resolve the close
pair $(3,1)$ than it is to resolve the less-close pairs
$(1,2)$ and $(2,3)$.  The weights (\ref{eq:wdefb}) appearing in
(\ref{eq:sbbar}) reflect the relative difficulty of the soft
radiation to resolve these different pairs.
Though I've been focused on the
single logs, the same is true of the double logs.  The $\ln y$
behavior in (\ref{eq:syint}), which
generates the double log after integration with $dy/y$,
also decomposes into contributions from
different pairs as
\begin {equation}
   \ln y = w_{12} \ln y + w_{23} \ln y + w_{31} \ln y .
\label {eq:lnyw}
\end {equation}

One may wonder why the weights $w_{ij}$ in (\ref{eq:sbbar}) and
(\ref{eq:lnyw}) care about the {\it relative} size of $\bar b_{ij}^2$
rather than the size of $\bar b_{ij}^2$ directly.  This is an
artifact of having factored out the full leading-order rate
$[d\Gamma/dx]_{\rm LO}$ in the definition (\ref{eq:syint}) of the soft
corrections.  Inserting a factor of $1 = w_{12}{+}w_{23}{+}w_{31}$ and
using (\ref{eq:bbarij}),
the formula
(\ref{eq:LOrate}) for $[d\Gamma/dx]_{\rm LO}$ can be
rewritten as
\begin {equation}
   \left[ \frac{d\Gamma}{dx} \right]_{\rm LO} =
   \frac{\alphas P(x)}{4\pi\sqrt2} \, e^{\gammaE} \qhatA
   (\bar b_{12}^2 + \bar b_{23}^2 + \bar b_{31}^2) .
\label {eq:altLOrate}
\end {equation}
Using this and (\ref{eq:dGij4}), or alternatively
returning to (\ref{eq:dGij2})
and directly using $b_{ij} = |x_k \bar B|$ and (\ref{eq:Bbar})
(and in either case also including the proper phase from
section \ref{sec:fix} for the amplitude-amplitude pair
$[d\Gamma/dx\,dy]_{12}$):
\begin {equation}
   \delta \left[ \frac{d\Gamma}{dx} \right] =
   \int_{y \ll x} dy \>
   \left[ \frac{d\Gamma}{dx\,dy} \right] ,
\label {eq:deldGdx}
\end {equation}
with
\begin {multline}
   \frac{d\Gamma}{dx\,dy} =
   \left[ \frac{d\Gamma}{dx\,dy} \right]_{12}
   + \left[ \frac{d\Gamma}{dx\,dy} \right]_{23}
   + \left[ \frac{d\Gamma}{dx\,dy} \right]_{31}
\\
   =
   - \frac{\CA \alphas^2 P(x)}{16\pi^2 y \sqrt2}\, e^{\gammaE} \qhatA
   \Bigl\{
     \bar b_{12}^2 \bigl[ \ln y + \bar s_\LMW(\bar b_{12}) \bigr]
     + \bar b_{23}^2
       \bigl[ \ln y + \bar s_\LMW( \bar b_{23} ) - \tfrac{\pi}{2} \bigr]
\\
     + \bar b_{31}^2
       \bigl[ \ln y + \bar s_\LMW( \bar b_{31} ) - \tfrac{\pi}{2} \bigr]
   \Bigr\} .
\end {multline}
Alternatively, one may define a complex-valued
\begin {equation}
   \beta_{ij} \equiv |x_k| \bar B    
\end {equation}
in terms of complex-valued $\bar B$ (\ref{eq:Bbar}).  Then
\begin {subequations}
\label {eq:nice}
\begin {multline}
   \frac{d\Gamma}{dx\,dy}
   =
   - \frac{\CA \alphas^2 P(x)}{16\pi^2 y}\, e^{\gammaE} \qhatA
   \Re
   \Bigl\{
     \beta_{12}^2 \bigl[ \ln y
         + \bar s_{\mbox{\scriptsize\rm blue-blue}}(\beta_{12}) \bigr]
     + \beta_{23}^2 \bigl[ \ln y + \bar s_\LMW( \beta_{23} ) \bigr]
\\
     + \beta_{31}^2 \bigl[ \ln y + \bar s_\LMW( \beta_{31} ) \bigr]
   \Bigr\} ,
\end {multline}
which writes the final answer in terms of just $\bar s_{\LMW}$
for red-blue pairs of lines and $\bar s_{\mbox{\scriptsize blue-blue}}$
for blue-blue pairs of lines, without additional, explicitly written
$\pi$ terms.  Or one may compactly write everything in this form
in terms of $\bar s_\LMW$ by noting
\begin {equation}
   \Re
   \Bigl\{
     \beta_{12}^2 \bigl[ \ln y
         + \bar s_{\mbox{\scriptsize\rm blue-blue}}(\beta_{12}) \bigr]
   \Bigr\}
   =
   \Re
   \Bigl\{
     \beta_{12}^2 \bigl[ \ln y
         + \bar s_\LMW\bigl(|\beta_{12}|\bigr) \bigr]
   \Bigr\} .
\end {equation}
\end {subequations}


\section {Conclusion}
\label {sec:conclusion}

We have seen that, at the ``microscopic'' level, using the $p_\perp$-broadening
effective value of $\hat q_{\rm eff}$ inside the BDMPS-Z calculation of
hard $g{\to}gg$ scattering correctly reproduces not only double logs
but also the subleading,
single-log soft radiative corrections to hard $g{\to}gg$ splitting.
Using (\ref{eq:altLOrate}), (\ref{eq:deldGdx}) and (\ref{eq:nice}),
we can also express the final, soft-radiation corrected splitting rate
in terms of the $p_\perp$-broadening effective
$\hat q_{\rm eff}(\Delta b) = \hat q + \delta\hat q_\LMW(\Delta b)$
from (\ref{eq:dqLMW}) as
\begin {subequations}
\label {eq:FINAL}
\begin {equation}
   \left[ \frac{d\Gamma}{dx} \right]_{\rm LO}
   +
   \delta \left[ \frac{d\Gamma}{dx} \right]
   =
   \frac{\alphas P(x)}{4\pi}\,
   \Re
   \left\{
     \hat\beta_{12}^2 \sqrt{ \frac{\qhatA^{\rm eff}(|\beta_{12}|)}{E} }
     + \hat\beta_{23}^2 \sqrt{ \frac{\qhatA^{\rm eff}\bigl(\beta_{23}\bigr)}{E} }
     + \hat\beta_{31}^2 \sqrt{ \frac{\qhatA^{\rm eff}\bigl(\beta_{31}\bigr)}{E} }
     \,
   \right\}
\label {eq:final}
\end {equation}
through first order in soft radiative corrections $\delta\hat q$,
with complex-valued transverse separation scales
$\beta_{ij}$ given in terms of dimensionless $\hat\beta_{ij}$ by
\begin {equation}
   \beta_{ij} \equiv \frac{e^{-\gammaE/2}}{(\qhatA E)^{1/4}} \, \hat\beta_{ij}
\end {equation}
and
\begin {equation}
   \begin{pmatrix}
      \hat\beta_{12} \\ \hat\beta_{23} \\ \hat\beta_{31}
   \end{pmatrix}
   \equiv
   (\qhatA E)^{1/4}
   \sqrt{ \frac{1}{M_0 \Omega_0} }
   \begin{pmatrix} |x_3| \\ |x_1| \\ |x_2| \end{pmatrix}
   =
   \frac{ e^{-i\pi/8} }
        {
          \bigl[ \tfrac12 x(1{-}x) (1{-}x{+}x^2) ]^{1/4}
        }
   \begin{pmatrix} 1 \\ 1{-}x \\ x \end{pmatrix} .
\label {eq:betadef2}
\end {equation}
\end {subequations}
Because of the complex phase in (\ref{eq:betadef2}), the
$\qhatA^{\rm eff}\bigl(|\beta_{12}|\bigr)$ is not the same
thing as $\qhatA^{\rm eff}(\beta_{12})$, breaking the symmetry
between the three gluons in (\ref{eq:final}).
That means, at the level of soft radiative corrections,
one can no longer ignore the difference between $\hat q^{\rm eff}$
effects for particle pairs with (i) both particles in the amplitude
vs.\ (ii) one particle each in the amplitude and conjugate amplitude.
For traditional
BDMPS-Z based calculations in the $\hat q$ approximation,
the possibility of such differences has always been ignored.

One of the side benefits of the result of this paper is that it provides
a highly non-trivial cross-check of the very involved
calculations of {\it hard}\/
radiative corrections to $g{\to}gg$ in
refs.\ \cite{2brem,seq,dimreg,QEDnf,qcd}, from which the limit of
soft-radiative corrections was extracted in ref.\ \cite{logs}.
In particular, the ``$\pi$ terms'' in those results, such as the
$\pi$ terms in (\ref{eq:sbar}), required a great deal of fussy work
to correctly choose branch cuts at intermediate stages of the calculation.
It is reassuring to see everything match up exactly with the much
simpler derivation here for the case of soft radiative corrections.

In this paper, I have used a sharp IR cut-off $y_{\rm cut} E$ on the softest
radiative gluon (allowing coverage of at most the red region of
fig.\ \ref{fig:LMWregion}), because that was the explicit calculation
\cite{logs} that was available to compare to.
Readers may wonder what would happen in a more complete calculation
that also included the gray region of fig.\ \ref{fig:LMWregion} and
correctly handled the breakdown of the $\hat q$ approximation at
$\Delta t \sim \tau_0$.
Similarly, one might want to include running of the $\alphas$ associated
with the soft splitting.
My personal expectation is
that the ``microscopic'' version of the universality of
$\hat q$ shown in this paper will continue to hold (at least in the
large-$\Nc$ limit).  That is, I expect that if one repeated the
BDMPS-Z derivation using the full $\hat q_\LMW(\Delta b)$ (now including
the gray region and/or running of $\alphas$),%
\footnote{ I should perhaps say an ``LMW-like'' $\hat q_\LMW(\Delta
  b)$ that includes the gray region.  Once the $\hat q$ approximation
  breaks down, the details of the calculation can depend on the
  details of the medium.  LMW handle the breakdown of the $\hat q$
  approximation with expressions involving distribution functions of
  gluons in the medium.  It is currently unclear to me, at least,
  exactly how these should be defined for my own favorite application,
  which is to quark-gluon plasmas.  } then one would obtain a result
that correctly incorporated soft radiative corrections to hard
splitting $g {\to} gg$.  I expect this because, as in
fig.\ \ref{fig:LMWregion}, the time scales for soft radiation (and
especially for the gray region) are small compared to the formation
time $t_{\rm form}(x)$, and so the soft gluon curly line in
fig.\ \ref{fig:lozenge} extends only for a relatively short time,
during which the hard-particle lines are approximately straight with
approximately constant separation, just like in the calculation of the
soft correction $\delta\hat q$ to transverse momentum broadening
depicted in fig.\ \ref{fig:LMWdiags} (or alternatively
fig.\ \ref{fig:LMWdiags12}).  I believe that the basis for the
calculation that I've made in this paper depends merely on this
(large-$\Nc$) factorization, and not on the exact details of the
formula for the soft correction.  It would be interesting to explore
to what extent the ``macroscopic'' final
formula (\ref{eq:FINAL}) might also be robust,
were one to use the full $\hat q_\LMW(\Delta b)$ to
redo the BDMPS-Z calculation.

Finally,
a very interesting unresolved question is whether the results of this paper
require the large-$\Nc$ approximation, or whether
``microscopic universality'' for sub-leading single logs
holds for finite $\Nc$ as well.


\acknowledgments

This work was supported, in part, by the U.S. Department
of Energy under Grant No.~DE-SC0007974.
My thanks to Shahin Iqbal and Tyler Gorda for their work with me
to find the single log results of ref.\ \cite{logs}, which motivated
the current study.
I also thank Simon Caron-Huot for a 2018 discussion of 
possible differences in $\hat q_{\rm eff}$ when both particles are in the
amplitude, as referenced in footnote \ref{foot:simon}.
Thanks also to Bronislav Zakharov for answering a question about
references.
Finally, I am grateful to the anonymous referee for saving me
from making an unjustified claim about bare $\hat q$.


\appendix

\section{Complex phases in LMW's derivation}
\label {app:fix}

LMW \cite{LMW} were always interested in the real part of their
diagrammatic results.  Here, I clarify some of the complex phases
in LMW's derivation for the sake of my section \ref{sec:fix}, where
the result is needed without taking the real part.

Like LMW's discussion in their main text, I will work here in the
large-$\Nc$ limit, even though their results for $\delta\hat q$ for
transverse momentum broadening are more general and do not ultimately depend on
this limit.  Similarly, even though I am interested in
soft radiative corrections to momentum broadening of {\it gluons},
here I will follow LMW and focus on corrections to momentum
broadening of quarks.  In the context of the large-$\Nc$ limit,
their quark $\hat q$ is related to gluon $\qhatA$ by
\begin {equation}
   \hat q = \frac{\qhatA}{2} \qquad \mbox{(large $\Nc$)} .
\end {equation}
These details do not matter.  If one did the same calculations for
hard particles in any color representation $R$, one would find the same
final formula for the relative correction $(\delta \hat q_R)/\hat q_R$.
By sticking to the case considered by LMW, however, the discussion will
be simpler, and I will be able to more easily compare
formulas.

Table \ref{tab:translate} gives a translation between LMW's notation
and the notation I have used in the main text.
To simplify direct comparison to their equations, I will mostly
use LMW's notation in this appendix.  One exception has to do
with the fact that part of LMW's calculation is set up so that
their propagator $G$ and complex frequency $\omega_0$ represent
a situation where the soft gluon is first emitted in the conjugate
amplitude --- what in this paper I would draw as a red gluon.
With my conventions, I am interested in handling the case where the
soft gluon is emitted in the amplitude --- a blue gluon.
The relevant propagator and complex frequency are then
$G^*$ and $\omega_0^*$ in LMW's notation,
which I will call ${\cal G}$ and $\Omega_{\rm s}$
(see table \ref{tab:translate}).
The subscript on $\Omega_{\rm s}$ stands for ``soft.''%
\footnote{
  My $\Omega_{\rm s}$ here is called $\Omega_y$ in ref.\ \cite{logs}.
}

\begin{table}[t]
\begin {center}
\begin{tabular}{cc}
\hline\hline
\noalign{\vskip 0.3em}
  LMW & this paper \\
\hline
  $\omega$  & $\omega_y {=} yE $ \\
  $x_\perp$  & $\Delta b$ \\
  $\B_\perp$ & $\b$ of the soft gluon \\
  $t$       & $\Delta t_y$ \\
  $\Nc$     & $\CA$ (= $2\CF$ for $\Nc{\to}\infty$)\\
  $\hat q$  & see text \\
  $\omega_0 {\equiv} \sqrt{i\hat q/\omega}$
            & $\Omega_{\rm s}^* {\equiv}
                \left( \sqrt{i\qhatA/2\omega_y} \right)^*
                = \sqrt{-i\qhatA/2\omega_y} $\\
  $G$       & ${\cal G}^*$ \\
  $G_0$     & ${\cal G}^*_{\rm vac}$ \\
  $l_0$     & $\tau_0$ \\
\noalign{\vskip 0.3em}
\hline\hline
\end{tabular}
\end {center}
\caption{
  \label{tab:translate}
  Translations between the notation of LMW \cite{LMW} and
  this paper.
}
\end{table}


\subsection {Setup}

LMW carry out different parts of their derivation in slightly
different ways.  In some parts [their calculation of single logs from
boundary (b)], they implicitly treat the soft gluon as what I would
call a red gluon (one emitted first from the conjugate amplitude).  In
other parts [their calculation of single logs from boundary (a)], their
formulas implicitly treat it as what I would call a blue gluon
(emitted first from the amplitude).  None of that matters to their
application, because they need to take $\Re(\cdots)$ at the end.  But
I additionally need the case where I do not take the real part.
So, we need to first review the general starting formula from whence their
single log contributions are extracted.  Here, I will briefly summarize
the origin of that formula
in the language I have used in this paper.  My starting formula will be
a very minor variation of LMW's.

Imagine computing the
soft radiative correction to the amplitude-amplitude (blue-blue)
potential, represented by fig.\ \ref{fig:LMWdiags12}.  As discussed
in the main text, taking $\Re(\cdots)$ of the result will then be
equivalent to LMW's calculation.

Without
radiative corrections, the light-like Wilson loop has length
dependence proportional to
\begin {equation}
   e^{-i V_0(x_\perp) L}
\end {equation}
where
\begin {equation}
   V_0(x_\perp) = -\tfrac{i}{4} \hat q x_\perp^2
\end {equation}
as in (\ref{eq:Vb}).  The first-order correction represented
by fig.\ \ref{fig:LMWdiags12} is
\begin {multline}
  \delta \left[ e^{-i V(x_\perp) L} \right] =
  - \CFgg
     \int_{\rm soft} \frac{d\omega}{(2\pi)2\omega}
     \int_0^L dz_2 \int_0^{z_2} dz_1 \> e^{-i V_0(x_\perp) \, (L{-}z_2)}
\\ \times
     \frac{ \grad_{B_{1\perp}}\!\cdot\grad_{B_{2\perp}} }{ \omega^2 } \,
     {\cal G}(\B_{2\perp},z_2;\B_{1\perp},z_1) e^{-i V_0(x_\perp) \, z_1}
     \Biggl|_{\B_{2\perp}=0}^{\B_{2\perp}=\x_\perp}
     \Biggl|_{\B_{1\perp}=0}^{\B_{1\perp}=\x_\perp}
     ,
\label {eq:dWilson1}
\end {multline}
where $z_1$ and $z_2$ are the $z$ coordinates (equivalently times) of the
first and second vertices.
Above, the two
factors of $e^{-i V_0(x_\perp) \cdots}$ represent the contributions to
the Wilson loop (i) from after $z_2$ and (ii) from before $z_1$.%
\footnote{
  Roughly speaking, the factorization of medium correlations into (i) $z<z_1$,
  (ii) $z_1 < z < z_2$, and (iii) $z_2<z$ is a consequence of
  high-energy formation lengths being large compared to the correlation
  length $l_0$ of the medium, so that medium correlations appear
  approximately instantaneous compared to the time scale of splitting
  processes.
  More specifically, it's because LMW's boundary (a) corresponds to separations
  $z_2{-}z_1 \gg l_0$.  [It's actually $z_2{-}z_1 \gtrsim l_0$, but
  the $z_2{-}z_1 \sim l_0$ end of boundary (a), by itself, is not
  log enhanced and
  does not contribute to single logs.]
}
The
$d\omega/[(2\pi) 2\omega]$ comes from the usual
relativistic phase space measure $d^3p/[(2\pi)^3 2\omega]$ for the
(approximately on-shell) soft gluon, the transverse $d^2p_\perp/(2\pi)^2$
part of which is absent because we are working in transverse position
space instead of transverse momentum space.
$\CF$ is the quark quadratic Casimir
\begin {equation}
   \CF = \frac{\Nc}{2} \qquad \mbox{(large $\Nc$)} .
\end {equation}
The
$-\CFgg \, \grad_{B_{1\perp}}\!\cdot\grad_{B_{2\perp}} / \omega^2$ comes from
the vertices $i g T_{\rm color}^a v\cdot A^a$
where the soft gluons attach to the Wilson lines, summed
over transverse polarizations of the (nearly-collinear) soft gluon.
The four different terms added/subtracted by the combinations
of values of $\B_{1\perp}$ and $\B_{2\perp}$ indicated at the end
of (\ref{eq:dWilson1}) represent the four diagrams in
fig.\ \ref{fig:LMWdiags12}.%
\footnote{
  For the first two diagrams in
  the first line of fig.\ \ref{fig:LMWdiags12},
  there is an extra minus sign compared to the self-energy diagrams.
  In the Wilson loop formulation, this
  may be described as arising from the fact that, going around the Wilson
  loop, the integration $\oint dx\cdot A$ follows the two light-like
  Wilson lines in opposite directions, so that there is a relative minus
  sign associated with the vertex factor for the backward-going line.
}
The ${\cal G}$ is the two-dimensional
quantum mechanics Green function for the
propagation of the soft gluon in the medium, in the $\hat q$ approximation,
analogous to (\ref{eq:HOprop}).
In the LMW case where the hard particles
are quarks rather than gluons, and where the quarks are taken to have
fixed transverse positions $0$ and $\x_\perp$ as above, the analog of
the 3-particle potential (\ref{eq:V3}) is
\begin {equation}
  V_{\rm (qgq)}(0,\b,\x_\perp)
  = - \tfrac{i}{8} \qhatA \bigl[ |\b|^2 + |\b-\x_\perp|^2 \bigr]
  = - \tfrac{i}{4} \hat q \bigl[ |\b|^2 + |\b-\x_\perp|^2 \bigr]
  \qquad \mbox{(large $\Nc$)} ,
\label {eq:Vqgq}
\end {equation}
in which $\b$ is the transverse position of the soft gluon.
The $b_{31}^2$ term of (\ref{eq:V3}) does not appear here because
in the large-$\Nc$ limit the quarks cannot directly interact when there is
a gluon between them.  
Up to conventions concerning complex conjugation,
(\ref{eq:Vqgq}) is the potential that appears in LMW eq.\ (6).
Complex conjugation arises because I consider
the case of a gluon emitted first from the amplitude, whereas LMW eq.\ (6)
implicitly refers to the case where the gluon is
instead first emitted from the conjugate amplitude.  As a result,
the explicit formula for my propagator ${\cal G}$ in (\ref{eq:dWilson1})
is the complex conjugate $G^*$ of their formula for their propagator $G$ in
LMW eq.\ (8).%
\footnote{
  To see the relation, complex conjugate LMW eq.\ (6)
  and then multiply both sides
  by $i$ to get a Schr\"odinger-like equation for my ${\cal G}=G^*$.
  Reading off the potential $V$ from that Schr\"odinger equation
  reproduces my (\ref{eq:Vqgq}).
}

In order to make contact with LMW's starting formula, rewrite
(\ref{eq:dWilson1}) above as
\begin {subequations}
\label {eq:dWilson1alt}
\begin {equation}
  \delta \left[ e^{-i V(x_\perp) L} \right] =
     \int_{\rm soft} \frac{d\omega}{\omega} \>
     {\cal N}(x_\perp,\omega)
\end {equation}
with
\begin {multline}
  {\cal N}(x_\perp,\omega) \equiv
  - \frac{\CFas}{\omega^2}
     \int_0^L dz_2 \int_0^{z_2} dz_1 \> e^{-i V_0(x_\perp) \, (L{-}z_2)}
\\ \times
     \grad_{B_{1\perp}}\!\cdot\grad_{B_{2\perp}}
     {\cal G}(\B_{2\perp},z_2;\B_{1\perp},z_1) e^{-i V_0(x_\perp) \, z_1}
     \Biggl|_{\B_{2\perp}=0}^{\B_{2\perp}=\x_\perp}
     \Biggl|_{\B_{1\perp}=0}^{\B_{1\perp}=\x_\perp}
     .
\end {multline}
\end {subequations}
This is related to the starting equation for $N(x_\perp,\omega)$
in LMW eq.\ (12) by
\begin {equation}
   N(x_\perp,\omega)
   =
   \Re\bigl[ {\cal N}(x_\perp,\omega) - {\cal N}_{\rm vac}(x_\perp,\omega) \bigr] ,
\end {equation}
where ${\cal N}_{\rm vac}$ is the vacuum version of ${\cal N}$, which
corresponds to the limit $\hat q{\to}0$.

Rewrite the $dz_2 \> dz_1$ integration as integration over $(z_1+z_2)/2$
and $t \equiv z_2-z_1$.  In the limit of $L$ large compared to
the soft gluon formation time, one may approximate the upper limit
on the $t$ integral as $\infty$ and use translation invariance to
approximate the integral over $(z_1+z_2)/2$ as $L$:
\begin {multline}
  \delta \left[ e^{-i V(x_\perp) L} \right] \simeq
  - \CFas L \, e^{-i V_0(x_\perp) L}
     \int_{\rm soft} \frac{d\omega}{\omega^3}
     \int_0^\infty dt \> e^{i V_0(x_\perp) \, t}
\\ \times
     \grad_{B_{1\perp}}\!\cdot\grad_{B_{2\perp}}
     {\cal G}(\B_{2\perp},t;\B_{1\perp},0)
     \Biggl|_{\B_{2\perp}=0}^{\B_{2\perp}=\x_\perp}
     \Biggl|_{\B_{1\perp}=0}^{\B_{1\perp}=\x_\perp}
     .
\label {eq:dWilson2}
\end {multline}
From (\ref{eq:dWilson2}) and the formal perturbative expansion
$e^{-i V(x_\perp) L} \simeq e^{-iV_0(x_\perp)L} (1 - i \delta V(x_\perp) \,L)
 = e^{-iV_0(x_\perp)L} (1 - \tfrac14 \delta\hat q \,x_\perp^2 L)$,
identify
\begin {align}
  \tfrac14 \,\delta\hat q \, x_\perp^2
  &\simeq
  \CFas
     \int \frac{d\omega}{\omega^3}
     \int_0^\infty dt \> e^{i V_0(x_\perp) \, t}
     \grad_{B_{1\perp}}\!\cdot\grad_{B_{2\perp}}
     {\cal G}(\B_{2\perp},t;\B_{1\perp},0)
     \Biggl|_{\B_{2\perp}=0}^{\B_{2\perp}=\x_\perp}
     \Biggl|_{\B_{1\perp}=0}^{\B_{1\perp}=\x_\perp}
\nonumber\\[6pt]
  &=
  \CFas
     \int \frac{d\omega}{\omega^3}
     \int_0^\infty dt \> e^{\hat q x_\perp^2 t/4} \,
     \grad_{B_{1\perp}}\!\cdot\grad_{B_{2\perp}}
     {\cal G}(\B_{2\perp},t;\B_{1\perp},0)
     \Biggl|_{\B_{2\perp}=0}^{\B_{2\perp}=\x_\perp}
     \Biggl|_{\B_{1\perp}=0}^{\B_{1\perp}=\x_\perp}
  .
\label {eq:dWilson3}
\end {align}
In order to make further contact with LMW, let me rewrite
(\ref{eq:dWilson3}) as
\begin {equation}
  \delta\hat q \, L
  \simeq
  - \grad_{x_\perp}^2
  \int \frac{d\omega}{\omega} \>
  \widetilde{\cal N}(x_\perp,\omega)
\label {eq:usetildeN}
\end {equation}
with
\begin {multline}
  \widetilde{\cal N}(x_\perp,\omega)
  \equiv
  - \frac{\CFas}{\omega^2} \, L
     \int_0^\infty dt \> e^{\hat q x_\perp^2 t/4} \,
     \grad_{B_{1\perp}}\!\cdot\grad_{B_{2\perp}}
         {\cal G}(\B_{2\perp},t;\B_{1\perp},0)
     \Biggl|_{\B_{2\perp}=0}^{\B_{2\perp}=\x_\perp}
     \Biggl|_{\B_{1\perp}=0}^{\B_{1\perp}=\x_\perp}
  .
\label {eq:tildeN}
\end {multline}
In the limit taken, $\widetilde{\cal N}$ is related to the $N$ of
LMW eq.\ (12)
by
\begin {align}
   N(x_\perp,\omega)
   &=
   \Re\left[
   e^{-i V_0(x_\perp) L} \widetilde{\cal N}(x_\perp,\omega)
      - \widetilde{\cal N}_{\rm vac}(x_\perp,\omega)
   \right]
\nonumber\\
   &=
   \Re\left[
      e^{-\qhatA x_\perp^2 L/4} \widetilde{\cal N}(x_\perp,\omega)
      - \widetilde{\cal N}_{\rm vac}(x_\perp,\omega)
   \right] .
\end {align}

LMW's use of vacuum subtraction is extremely convenient computationally
but inessential:
As I briefly discuss in appendix \ref{sec:Nvac},
$\widetilde{\cal N}_{\rm vac}$ vanishes.
I will take advantage of this to be a little sloppy in what follows.
The reader may assume that later formulas are vacuum subtracted unless
I specifically refer to the vacuum piece.


\subsection {Crossing the boundary (b)}

LMW's boundary (b) refers to the upper boundary in my
fig.\ \ref{fig:LMWregion}.
In LMW eq.\ (27), they write
\begin {equation}
   N(x_\perp,\omega) =
   i \frac{\alphas\Nc x_\perp^2\omega}{4\pi}
   \int_{t_0}^L dt \> \frac{L-t}{t^3}
   \left\{
     \left( \frac{\omega_0 t}{\sin(\omega_0 t)} \right)^{\!\!3}
       \bigl[4 - \sin^2(\omega_0 t) \bigr]
     - 4
   \right\} .
\label {eq:LMWboundaryb1}
\end {equation}
There is an implicit real part $\Re(\cdots)$ of the right-hand side
of the equation [see LMW eq. (19)], which they did not write explicitly.
$t_0$ is an ($\omega$ dependent) scale chosen to lie between the two
boundaries, parametrically far from either:
\begin {equation}
   \frac{\omega}{\hat q L} \ll t_0 \ll \frac{1}{|\omega_0|} .
\label {eq:t0param}
\end {equation}

Taking small-$x_\perp$ limits appropriate to boundary (b) the same way as
in LMW, I find that my $\widetilde{\cal N}$ of (\ref{eq:tildeN}) is given by
the complex conjugate of the right-hand side of (\ref{eq:LMWboundaryb1}).
In the relevant large-$L$ limit $L \gg 1/|\omega_0|$,
\begin {equation}
   \widetilde{\cal N}(x_\perp,\omega) \simeq
   -i \frac{\alphas\Nc x_\perp^2\omega}{4\pi} \, L
   \int_{t_0}^\infty \frac{dt}{t^3}
   \left\{
     \left( \frac{\Omega_{\rm s} t}{\sin(\Omega_{\rm s} t)} \right)^{\!\!3}
       \bigl[4 - \sin^2(\Omega_{\rm s} t) \bigr]
     - 4
   \right\} ,
\label {eq:LMwboundaryb1}
\end {equation}
where I define $\Omega_{\rm s} \equiv \omega_0^*$ as described earlier.

In the $|\Omega_s| t_0 \ll 1$ limit of (\ref{eq:t0param}), the integral
gives
\begin {equation}
   \widetilde{\cal N}(x_\perp,\omega) \simeq
   \frac{\alphas \Nc x_\perp^2}{4\pi} \, \hat q L
   \left[ \ln\left( \frac{i \Omega_{\rm s} t_0}{2} \right) + \frac13 \right] .
\label {eq:tildeNb}
\end {equation}
If one takes the real part, this agrees with the LMW eq.\ (28) result that
\begin {equation}
   N(x_\perp,\omega) \simeq
   \frac{\alphas \Nc x_\perp^2}{4\pi} \, \hat q L
   \Re \left[ \ln\left( \frac{\omega_0 t_0}{2} \right) + \frac13 \right] ,
\end {equation}
where I have here made explicit the implicit $\Re(\cdots)$ in LMW's equation.%
\footnote{
  I've also fixed a trivial typographic error by restoring a missing
  factor of $\alphas$.
}

Using (\ref{eq:tildeNb}) in (\ref{eq:usetildeN}), the contribution to
$\hat q$ from LMW's boundary (b) is
\begin {align}
  \delta\hat q \big|_{\rm boundary~b}
  &= \hat q \int_{\rm soft} \frac{d\omega}{\omega} \>
    \frac{\alphas\Nc}{\pi}
    \left[ \ln\left( \frac{2}{i\Omega_{\rm s} t_0} \right) - \frac13 \right]
\nonumber\\
  &= \hat q \int_{\rm soft} \frac{d\omega}{\omega} \>
    \frac{\alphas\Nc}{\pi}
    \left[
       \ln\left(
          \frac{2}{t_0} \sqrt{ \frac{\omega}{\hat q} }
          \, e^{-i\pi/4}
       \right)
       - \frac13
  \right] .
\label {eq:boundaryb}
\end {align}
If the real part is taken, as would be appropriate in the case of
red-blue pairs of lines, this result becomes equivalent to
LMW eq.\ (29).%
\footnote{
  Specifically, divide both sides of LMW eq.\ (29) by $\omega L$ and
  then integrate over soft $\omega$.
}


\subsection {Crossing the boundary (a)}

LMW's boundary (a) refers to the lower boundary of the red region in my
fig.\ \ref{fig:LMWregion}.  Parametrically, it corresponds to
\begin {equation}
  t \sim \frac{\omega}{\hat q L} \ll \frac{1}{|\omega_0|} \,.
\label {eq:tboundarya}
\end {equation}
Expand the general formula (\ref{eq:tildeN}) for $\widetilde{N}$ in
small $|\omega_0| t$ and also small $x_\perp$, making no assumption
about the size of $\omega x_\perp^2/t$.
The leading order result (after an implicit vacuum subtraction) is
\begin {equation}
  \widetilde{\cal N}(x_\perp,\omega)
  \simeq
  -\frac{\alphas\Nc}{12\pi} \, \hat q L \int_0^{t_0} dt \>
   \left[
     \left( \frac{x_\perp^2}{t} + \frac{i\omega(x_\perp^2)^2}{2t^2} \right)
          e^{i\omega x_\perp^2/2t}
     + \frac{4i}{\omega} (1 - e^{i\omega x_\perp^2/2t} )
   \right] ,
\label {eq:tildeNa1}
\end {equation}
where $t$ is integrated over $t < t_0$ because here we are focused
on the range of $t$ that contains boundary (a) instead of boundary (b).
For the sake of contact with the discussion in LMW,
I should mention that
(\ref{eq:tildeNa1}) turns out to be equivalent to
an expansion of the general formula (\ref{eq:tildeN}) for
$\widetilde{\cal N}$ to first order in
$\hat q$ (remembering that the definition of $\omega_0$ depends on $\hat q$).
The real part of (\ref{eq:tildeNa1}) is the same as LMW eq.\ (32).

Doing the time integral in (\ref{eq:tildeNa1}) gives
\begin {equation}
   \widetilde{\cal N}(x_\perp,\omega) \simeq
   \frac{\alphas\Nc x_\perp^2}{4\pi} \,
     \hat q L
     \left[
       \ln\left( -\frac{i\omega x_\perp^2}{2t_0} \right) + \gammaE - \frac13
     \right] ,
\label {eq:tildeNa}
\end {equation}
whose real part corresponds to LMW eq.\ (33) [which has an implicit
$\Re(\cdots)$].
Using (\ref{eq:tildeNa}) in (\ref{eq:usetildeN}), the contribution to
$\hat q$ from LMW's boundary (a) is
\begin {align}
  \delta\hat q \big|_{\rm boundary~a}
  &= -\hat q \int_{\rm soft} \frac{d\omega}{\omega} \>
    \frac{\alphas\Nc}{\pi}
    \left[
       \ln\left( -\frac{i\omega x_\perp^2}{2 t_0} \right)
       + \gammaE
       - \frac13
    \right]
\nonumber\\
  &= \hat q \int_{\rm soft} \frac{d\omega}{\omega} \>
    \frac{\alphas\Nc}{\pi}
    \left[
       \ln\left( \frac{2 t_0}{\omega x_\perp^2} \, e^{i\pi/2} \right)
       + \frac13
       - \gammaE
    \right]
\label {eq:boundarya}
\end {align}
The real part is equivalent to LMW eq.\ (34).


\subsection {Total}

In this paper, I am focused only on what LMW call boundaries (a) and (b).
Adding (\ref{eq:boundaryb}) and (\ref{eq:boundarya}) gives
\begin {align}
  \delta\hat q
  &= \hat q \left[
      -\frac{\alphas\Nc}{\pi}
      \int_{\rm soft} \frac{d\omega}{\omega} \>
      \left[
       \ln\left( \tfrac14 x_\perp^2 \Omega_{\rm s} \omega \right)
       + \gammaE
      \right]
    \right]
\nonumber\\
  &= \hat q \left[
      -\frac{\alphas\Nc}{\pi}
      \int_{\rm soft} \frac{d\omega}{\omega} \>
      \left[
       \ln\left( \tfrac14 x_\perp^2 \sqrt{\hat q \omega} \, e^{-i\pi/4} \right)
       + \gammaE
      \right]
    \right] .
\end {align}
Taking the real part and translating to my notation gives
the LMW result in the form I quoted in (\ref{eq:LMW}).
Not taking the real part gives my (\ref{eq:sblueblue}) instead.


\subsection {A brief word about $\widetilde{\cal N}_{\rm vac}$}
\label{sec:Nvac}

Earlier, I mentioned that the vacuum contribution
\begin {equation}
  \widetilde{\cal N}_{\rm vac}(x_\perp,\omega)
  =
  - \frac{\CF\kern0.6pt\alphas}{\omega^2} \, L
     \int_0^\infty dt \>
     \grad_{B_{1\perp}}\!\cdot\grad_{B_{2\perp}}
       {\cal G}_{\rm vac}(\B_{2\perp},t;\B_{1\perp},0))
     \Biggl|_{\B_{2\perp}=0}^{\B_{2\perp}=\x_\perp}
     \Biggl|_{\B_{1\perp}=0}^{\B_{1\perp}=\x_\perp}
\label {eq:Nvac}
\end {equation}
vanishes.
In the case of the LMW application to
momentum broadening, this is because an on-shell hard particle
cannot radiate in vacuum, because of energy-momentum conservation.
The large-$L$ limit that we took to get to (\ref{eq:tildeN})
formally disposed of any vacuum radiation associated with the particle
being ``created'' or ``destroyed'' at the ends of the lightlike
Wilson lines (i.e.\ at $z=0$ or $z=L$).
So $\Re(\widetilde N_{\rm vac})$ must vanish, and it turns out that
the imaginary part vanishes as well.
Showing this mathematically from (\ref{eq:Nvac})
is a little tricky because of divergences associated
with $t{\to}0$, which I will now briefly discuss and resolve.

Eq.\ (\ref{eq:Nvac}) corresponds
to setting $\hat q$ to zero and using the vacuum version
\begin {equation}
  {\cal G}_{\rm vac}(\B_{2\perp},t;\B_{1\perp},0)
  =
  \frac{\omega}{2\pi i t} \, e^{i\omega |\B_{2\perp}-\B_{1\perp}|^2 / 2t}
\label {eq:Gvac}
\end {equation}
of the propagator ${\cal G}$ in (\ref{eq:tildeN}).
That gives
\begin {equation}
  \widetilde{\cal N}_{\rm vac}(x_\perp,\omega)
  =
  \frac{\CFas}{\pi} \, L \int_0^\infty dt \>
   \left[
     - \frac{i\omega x_\perp^2}{t^3} e^{i\omega x_\perp^2/2t}
     + \frac{2}{t^2} \bigl(1 - e^{i\omega x_\perp^2/2t} \bigr)
   \right] .
\label {eq:Nvac2}
\end {equation}
The rapid oscillation of $e^{i\omega x_\perp^2/2t}$ as $t\to0$ makes
those terms in the integrand have a convergent integral.
However, the integral of $2/t^2$ in (\ref{eq:Nvac2}) has no such
convergence factor, and so the integral, as written, is ill-defined.

One may be able to argue correct $i\epsilon$ prescriptions
for the $t$ integral.  However, in my experience \cite{dimreg},
it is often less confusing to handle $t{\to}0$ divergences in LPM effect
calculations by using dimensional regularization.

Let $d \equiv d_\perp = 2-\eps$ be the number of transverse dimensions.
The $d$-dimensional version of the vacuum propagator
(\ref{eq:Gvac}) is
\begin {equation}
  {\cal G}_{\rm vac}(\B_{2\perp},t;\B_{1\perp},0)
  =
  \left( \frac{\omega}{2\pi i t} \right)^{d/2}
  e^{i\omega |\B_{2\perp}-\B_{1\perp}|^2 / 2t}
\label {eq:Gvacd}
\end {equation}
Then (\ref{eq:Nvac}) gives
\begin {equation}
  \widetilde{\cal N}_{\rm vac}(x_\perp,\omega)
  \propto
  \int_0^\infty dt \>
   \left[
     - \frac{i\omega x_\perp^2}{t^{2+\frac12 d}} \, e^{i\omega x_\perp^2/2t}
     + \frac{d}{t^{1+\frac12 d}} \, \bigl(1 - e^{i\omega x_\perp^2/2t} \bigr)
   \right] ,
\label {eq:Nvacd}
\end {equation}
which for $d{=}2$ is the time integral in (\ref{eq:Nvac2}).
But in dimensional regularization we may analytically continue the
result from values of $d$ where it converges ($-2 < d < 0$) and is unambiguous.
The integrand in (\ref{eq:Nvacd}) is a total derivative:%
\begin {equation}
  \widetilde{\cal N}_{\rm vac}(x_\perp,\omega)
  \propto
  \int_0^\infty dt \>
     \frac{d}{dt}
   \left[
     -\frac{2}{t^{d/2}} \, \bigl(1 - e^{i\omega x_\perp^2/2t} \bigr)
   \right] = 0 .
\end {equation}


\end {document}